\documentstyle[12pt]{article}
\makeatletter

\expandafter\ifx\csname mathrm\endcsname\relax\def\mathrm#1{{\rm #1}}\fi

\makeatother

\def\beq{\begin{equation}}
\def\eeq{\end{equation}}
\def\beqar{\begin{eqnarray}}
\def\eeqar{\end{eqnarray}}
\def\barr#1{\begin{array}{#1}}
\def\earr{\end{array}}
\def\bfi{\begin{figure}}
\def\efi{\end{figure}}
\def\btab{\begin{table}}
\def\etab{\end{table}}
\def\bce{\begin{center}}
\def\ece{\end{center}}
\def\nn{\nonumber}
\def\disp{\displaystyle}
\def\text{\textstyle}

\def\ga{\gamma}
\def\Ga{\Gamma}
\def\de{\delta}
\def\De{\Delta}

\def\la{\lambda}
\def\si{\sigma}
\def\Si{\Sigma}

\def\om{\omega}
\def\omp{\om_{+}}
\def\omm{\om_{-}}

\def\refeq#1{\mbox{(\ref{#1})}}
\def\reffi#1{\mbox{Fig.~\ref{#1}}}
\def\reffis#1{\mbox{Figs.~\ref{#1}}}

\def\refse#1{\mbox{Sect.~\ref{#1}}}
\def\refsse#1{\mbox{Subsect.~\ref{#1}}}

\def\refapp#1{\mbox{App.~\ref{#1}}}
\def\citere#1{\mbox{Ref.~\cite{#1}}}
\def\citeres#1{\mbox{Refs.~\cite{#1}}}

\renewcommand{\O}{{\cal O}}

\def\mathswitchr#1{\relax\ifmmode{\mathrm{#1}}\else$\mathrm{#1}$\fi}

\newcommand{\PW}{\mathswitchr W}
\newcommand{\PZ}{\mathswitchr Z}

\newcommand{\Pg}{\mathswitchr g}
\newcommand{\PH}{\mathswitchr H}

\newcommand{\Pf}{\mathswitchr f}
\newcommand{\Pfbar}{\mathswitchr {\bar f}}

\newcommand{\Pb}{\mathswitchr b}
\newcommand{\Pt}{\mathswitchr t}
\newcommand{\Ptbar}{\mathswitchr {\bar t}}

\newcommand{\PWp}{\mathswitchr {W^+}}
\newcommand{\PWm}{\mathswitchr {W^-}}

\def\mathswitch#1{\relax\ifmmode#1\else$#1$\fi}

\newcommand{\mfi}{\mathswitch {m_{{\mathrm f}_i}}}
\newcommand{\Mfp}{\mathswitch {M_{\Pf^\prime}}}
\newcommand{\Mf}{\mathswitch {M_\Pf}}

\newcommand{\MW}{\mathswitch {M_\PW}}

\newcommand{\MZ}{\mathswitch {M_\PZ}}
\newcommand{\MH}{\mathswitch {M_\PH}}

\newcommand{\Mb}{\mathswitch {m_\Pb}}
\newcommand{\Mt}{\mathswitch {m_\Pt}}

\newcommand{\sw}{\mathswitch {s_{\ss\PW}}}
\newcommand{\cw}{\mathswitch {c_{\ss\PW}}}

\newcommand{\D}{\displaystyle}
\newcommand{\Scr}{\scriptstyle}
\newlength{\zeichen}
\newlength{\strich}
\newlength{\sstrich}
\newcommand{\Slash}[1]
{#1\mbox{\settowidth{\zeichen}{$\D #1 $}\hspace{-0.5\zeichen}%
\hspace{-0.5\strich}}/}
\newcommand{\sSlash}[1]
{#1\mbox{\settowidth{\zeichen}{$\Scr #1 $}\hspace{-0.5\zeichen}%
\hspace{-0.5\sstrich}}/}

\newcommand{\phrd}[1]{Phys.\ Rev.\ {\bf D#1}}
\newcommand{\phr}[1]{Phys.\ Rev.\ {\bf #1}}
\newcommand{\phrl}[1]{Phys.\ Rev.\ Lett.\ {\bf #1}}
\newcommand{\nphb}[1]{Nucl.\ Phys.\ {\bf B#1}}
\newcommand{\nphbps}[1]{Nucl.\ Phys.\ {\bf B} (Proc.\ Suppl.) {\bf #1B}}

\newcommand{\phlb}[1]{Phys.\ Lett.\ {\bf B#1}}

\newcommand{\zphc}[1]{Z.\ Phys.\ {\bf C#1}}
\newcommand{\aph}[1]{Ann.\ Phys.\ (NY) {\bf #1}}

\newcommand{\phrp}[1]{Phys.\ Rep.\ {\bf #1}}

\newcommand{\fp}[1]{Fortschr.\ Phys.\ {\bf #1}}

\def\tr#1{\,\mbox{tr}\left\{#1\right\}}

\newcommand{\spc}{\phantom{{}+{}}}

\newcommand{\lag}{{\cal {L}}}
\newcommand{\leff}{\lag_{\mathrm{eff}}}
\newcommand{\lreff}{\lag_{\mathrm{eff}}^{\mathrm{ren}}}
\newcommand{\lcth}{\lag_{\mathrm{\hH}}^{\mathrm{ct}}}

\newcommand{\DW}{{\hat{D}_W}}
\newcommand{\OW}{{\overline{W}}}

\newcommand{\SY}{S_{\mathrm{Y}}}
\newcommand{\vp}{\varphi}
\newcommand{\hfi}{\hat{\mathrm{f}}_i}
\newcommand{\hfibar}{\hat{\bar{\mathrm{f}}}_i}

\newcommand{\hvp}{\hat{\varphi}}
\newcommand{\hH}{\hat{H}}
\newcommand{\hW}{\hat{W}}
\newcommand{\hZ}{\hat{Z}}
\newcommand{\hA}{\hat{A}}
\newcommand{\hB}{\hat{B}}
\newcommand{\hC}{\hat{C}}
\newcommand{\hD}{\hat{D}}
\newcommand{\hT}{\hat{T}}
\newcommand{\hU}{\hat{U}}
\newcommand{\hUd}{\hat{U}^\dagger}
\newcommand{\hV}{\hat{V}}
\newcommand{\Psif}{\Psi_{f}}
\newcommand{\Psibf}{\overline{\Psi}_{f}}
\newcommand{\hp}{\hat{\Psi}_f}
\newcommand{\hpb}{\hat{\overline{\Psi}}_f}
\newcommand{\hpp}{\hat{\Psi}_{f^\prime}}
\newcommand{\hpbp}{\hat{\overline{\Psi}}_{f^\prime}}
\newcommand{\hF}{\hat{F}}
\newcommand{\bX}{\overline{X}}
\renewcommand{\ss}{\scriptscriptstyle}
\newcommand{\intdp}{\int\frac{d^4 p}{(2\pi)^4}\,}
\renewcommand{\c}{\zeta}
\newcommand{\ord}[1]{\O(\c^{#1})}
\newcommand{\OH}[1]{\O(\MH^{#1})}
\newcommand{\dpp}{(x,\partial_x+i p)}
\newcommand{\tth}{ \tilde{\tilde{\Delta}}_H\dpp}
\newcommand{\tthn}{ \tilde{\tilde{\Delta}}_H}
\newcommand{\nl}{\nn\\&&{}}
\newcommand{\tde}{\De_{\MH}}
\newcommand{\fac}{\displaystyle\frac{1}{16\pi^2}}
\newcommand{\back}{\!\!\!\!\!\!\!\!\!\!\!\!}
\newcommand{\toug}{\quad\stackrel{\mathrm{U-gauge}}{\longrightarrow}\quad}

\def\Re{\mathop{\mathrm{Re}}\nolimits}

\def\draftdate{\relax}
\def\mda{\relax}
\def\mua{\relax}
\def\mla{\relax}
\def\draft{
\def\thtystars{******************************}
\def\sixtystars{\thtystars\thtystars}
\typeout{}
\typeout{\sixtystars**}
\typeout{* Draft mode!
         For final version remove \protect\draft\space in source file *}
\typeout{\sixtystars**}
\typeout{}
\def\draftdate{\today}
\def\mua{\marginpar[\boldmath\hfil$\uparrow$]%
                   {\boldmath$\uparrow$\hfil}%
                    \typeout{marginpar: $\uparrow$}\ignorespaces}
\def\mda{\marginpar[\boldmath\hfil$\downarrow$]%
                   {\boldmath$\downarrow$\hfil}%
                    \typeout{marginpar: $\downarrow$}\ignorespaces}
\def\mla{\marginpar[\boldmath\hfil$\rightarrow$]%
                   {\boldmath$\leftarrow $\hfil}%
                    \typeout{marginpar: $\leftrightarrow$}\ignorespaces}
\overfullrule 5pt
\oddsidemargin -15mm
\marginparwidth 29mm
}

\def\stars{\strut\leaders\hbox{*}\hfill\strut}
\def\starline{\hfil\strut\hfil\hbox to \textwidth {\stars}\hfil}

\renewcommand{\theequation}{\thesection.\arabic{equation}}
\newcounter{saveeqn}

\makeatletter
\@addtoreset{equation}{section}

\def\eqnarray{\stepcounter{equation}\let\@currentlabel=\theequation
\global\@eqnswtrue
\global\@eqcnt\z@\tabskip\@centering\let\\=\@eqncr
$$\halign to \displaywidth\bgroup\hskip\@centering
  $\displaystyle\tabskip\z@{##}$\@eqnsel&\global\@eqcnt\@ne
  \hskip 2\arraycolsep \hfil${##}$\hfil
  &\global\@eqcnt\tw@ \hskip 2\arraycolsep $\displaystyle\tabskip\z@{##}$\hfil
   \tabskip\@centering&\llap{##}\tabskip\z@\cr}
\def\appendix{\par
 \setcounter{section}{0} \setcounter{subsection}{0}
 \def\thesection{\Alph{section}}}

\makeatother

\begin{document}

\begin{titlepage}
\title{Integrating out the Standard Higgs Field \\
in the Path Integral}
\author{Stefan Dittmaier${}^1$\thanks{
E-Mail: dittmair@physik.uni-bielefeld.de}~\thanks{
Partially supported by the Bundesministerium f\"ur
Bildung und
Forschung,
Bonn, Germany.}
{}~and Carsten Grosse-Knetter${}^{1,2}$
\\[5mm]
\normalsize ${}^1$Universit\"at Bielefeld, Fakult\"at f\"ur Physik,\\
\normalsize Postfach 10 01 31, D-33501 Bielefeld, Germany\\[5mm]
\normalsize ${}^2$Universit\'e de Montr\'eal, Laboratoire de Physique
Nucl\'eaire,\\
\normalsize C.P. 6128, Montr\'eal, Qu\'ebec, H3C 3J7, Canada}
\date{BI-TP 95/10\\UdeM-GPP-TH-95-21\\
hep-ph/9505266\\
May 1995}
\maketitle
\thispagestyle{empty}

\begin{abstract}
We integrate out the Higgs boson in the electroweak standard model
at one loop and construct a low-energy effective Lagrangian
assuming that the Higgs mass is much larger than the
gauge-boson masses.
Instead of
applying diagrammatical techniques, we integrate out the Higgs boson
directly in the path integral, which turns out to be
much simpler. By using the background-field method and the
Stueckelberg formalism, we directly find a manifestly gauge-invariant
result. The heavy-Higgs effects on fermionic couplings are derived, too.
At one loop the $\log\MH$-terms of the
heavy-Higgs limit of the electroweak standard model coincide with
the UV-divergent terms in the gauged non-linear $\sigma$-model, but
vertex functions differ in addition by finite
constant terms.
Finally, the leading Higgs effects to some physical processes
are calculated from the effective Lagrangian.
\end{abstract}
\end{titlepage}

\section{Introduction}

In a previous article \cite{sdcgk} we have developed a method to
eliminate non-decoupling
heavy particles from a theory and to construct a one-loop effective Lagrangian
which parametrizes the low-energy  effects of these heavy particles.
We have applied
functional methods, i.e.\ instead of
calculating the effects of the the heavy
fields diagrammatically, we have integrated
them out directly in the path
intgral.
The contributions of the generated functional
determinant to the effective Lagrangian
have been expanded
in inverse powers of the heavy
mass. In \citere{sdcgk} this method has been explained
in detail by considering
a simple toy model, viz. by integrating out the heavy Higgs boson in
an SU(2) gauged linear $\sigma$-model
without fermions.

In the present article we apply this method to a phenomenologically
interesting  example:
we consider the $\mathrm{SU(2)_{\mathrm{W}}\times U(1)_{\mathrm{Y}}}$
electroweak standard model (SM) and assume that the
Higgs boson has a large mass in comparison to the gauge-boson
and fermion masses and the external momenta of the
scattering processes under consideration. We
integrate out the Higgs boson and determine its non-decoupling
effects, i.e.\
we calculate the $\OH{0}$-terms (which includes the $\log\MH$-terms)
of the corresponding low-energy
effective Lagrangian,
including the effective terms with fermion fields.
This way we formally construct the limit
$\MH\to\infty$ of the SM at one loop, which is a good approximation to
the physically interesting case of a finite but heavy Higgs mass
close to the unitarity limit of $\MH\sim 1$ TeV.
The leading one-loop Higgs contributions to scattering processes
and physical parameters can then easily
be derived from the effective Lagrangian. This will be discussed
by considering some examples.

Our method to integrate out  heavy fields in the path integral
has been discussed in detail in \citere{sdcgk}.
Therefore, we will present all those parts of our calculation
only very briefly which concern this method in general
or which can be done in analogy to the SU(2) model without fermions
considered in \citere{sdcgk}.
Different methods to construct low-energy effective Lagrangians by integrating
out heavy fields have been proposed in \cite{gale,fume,chan,chey}.

The Higgs boson has recently been integrated out in the
SM without fermions by diagrammatic methods in \citere{hemo}.
The result of our functional calculation
agrees with the
one given there.
Comparing our
functional calculation with the diagrammatic one,
we find that the functional method simplifies the calculation very
much. While in a diagrammatic calculation one has to calculate
the Higgs-dependent contributions to various Green functions
(i.e.\ very many Feynman graphs)
and then determine the coupling constants
of the effective Lagrangian by comparing coefficients
(``matching''), in a
functional calculation the effective Lagrangian is generated
{\em directly.\/} For instance,
there are 14 effective bosonic interaction terms
which are expected to be generated by naive power counting.
In fact only 7 of these terms are generated,
but the others (viz. the custodial
SU(2)$_\PW$-violating dimension-4 terms) are not.
In a diagrammatic calculation one has first
to consider all these terms when comparing the coefficients, and then it
turns out that they vanish.
However, in a functional calculation it is obvious that they
are suppressed by at least a factor $\MW^2/\MH^2$.
The use of the background-field method \cite{bfm1,bfm2,bfm3,bfm4,bfm5}
and the Stueckelberg Formalism \cite{stue1,stue2,stue4,stue3}
automatically ensures the gauge
invariance of the generated effective terms, while in
the conventional formalism there are some subleties
concerning gauge invariance of the
matching conditions \cite{esma}.

In addition to the treatment of the bosonic sector of the SM,
we also determine the effects
of a heavy Higgs boson on fermionic interactions,
which have not been calculated before.
All effective fermionic
interactions are proportional to $m_{\mathrm f}/\MW$ and thus
suppressed for all fermions except for the top quark.

This article is organized as follows: In \refse{sec:bfmstf} we describe the
background-field method and the Stueckelberg formalism for the
bosonic part of the electroweak standard model and determine the
one-loop part of the Lagrangian.
In \refse{sec:diag} we diagonalize the Higgs part of this Lagrangian.
In \refse{sec:helim} we integrate out the quantum Higgs field and
construct the effective Lagrangian, which
is written in a manifestly gauge-invariant standard form
in \refse{sec:invstue}.
In \refse{sec:ren} we carry out the renormalization
of the Higgs sector.
In \refse{sec:bghelim} the background Higgs field is
eliminated, which yields the final effective Lagrangian.
In \refse{sec:fermions} we integrate out the Higgs boson in the
fermionic part of the SM and calculate the fermionic terms of the
effective Lagrangian.
Section~\ref{sec:dis} contains the discussion of the result.
In \refse{sec:examples} we derive the $\log\MH$-contributions
to some physical processes directly from our effective Lagrangian.
Section~\ref{sec:sum} contains  our conclusions.
In \refapp{app:ints} the explicit form of
the Feynman integrals occurring in the calculations are given.
In \refapp{app:Pind} we prove an identity
needed for our calculation.

\section{The background-field method and the Stueckelberg formalism}
\label{sec:bfmstf}

\subsection{The standard-model Lagrangian}
\label{ssec:smlag}

In this and the subsequent sections
we first consider only the bosonic sector of the
$\mathrm{SU(2)_{\mathrm W}\times U(1)_{\mathrm Y}}$ electroweak SM.
The fermions will be included in \refse{sec:fermions}.
The bosonic part of the SM is specified by the Lagrangian
\beqar
\lag &=& -\frac{1}{2} \tr{W_{\mu\nu}W^{\mu\nu}}
-\frac{1}{4}B_{\mu\nu}B^{\mu\nu} \nn\\
&& +\frac{1}{2}\tr{(D_\mu\Phi)^\dagger(D^\mu\Phi)}
+\frac{1}{2}\mu^2\tr{\Phi^\dagger\Phi}
-\frac{1}{16}\lambda\left(\tr{\Phi^\dagger\Phi}\right)^2.
\label{eq:smlag}
\eeqar
The field-strength tensors
$W^{\mu\nu}$ and $B^{\mu\nu}$
read
\beqar
W^{\mu\nu}&=&\partial^\mu W^\nu-\partial^\nu
W^\mu-ig_2[W^\mu,W^\nu],\nn\\
B^{\mu\nu}&=&\partial^\mu B^\nu-\partial^\nu B^\mu,
\label{eq:fst}
\eeqar
where $W^\mu=W^\mu_i\tau_i/2$ and $B^\mu$ represent the corresponding
gauge fields. We note that we use the convenient matrix notation for the
$\mathrm{SU(2)_{\mathrm W}}$
representations throughout, with $\tau_i$ denoting the
Pauli matrices. The covariant derivative $D^\mu\Phi$ of the scalar Higgs
doublet $\Phi$ is given by
\beq
D^\mu\Phi = \partial^\mu\Phi-ig_2W^\mu\Phi-ig_1\Phi B^\mu\frac{\tau_3}{2}.
\label{eq:dphi}
\eeq
Usually, the field $\Phi$ is linearly represented by
\beq
\Phi=\frac{1}{\sqrt{2}}\left((v+H){\bf 1}+2i\vp\right),
\label{eq:philin}
\eeq
where $H$ is the (physical) Higgs field and $\vp=\vp_i\tau_i/2$ the
(unphysical) Goldstone field. The non-vanishing vacuum expectation value
is quantified by
\beq
v=2\sqrt{\frac{\mu^2}{\lambda}}.
\label{eq:vev}
\eeq
For our purpose it is much more appropriate to use the following
non-linear representation
\beq
\Phi=\frac{1}{\sqrt{2}}(v+H)U \qquad\mbox{with}\quad
U=\exp\left(2i\frac{\varphi}{v}\right),
\label{eq:phinl}
\eeq
where $H$ is an $\mathrm{SU(2)_{\mathrm W}}$ singlet, and the Goldstone fields
$\vp_i$ form the unitary matrix $U$.
In both representations the charge eigenstates of $\vp$ are given by
\beq
\vp^\pm = \frac{1}{\sqrt{2}}\left(\vp_2\pm i\vp_1\right),
\qquad
\chi = -\vp_3.
\eeq
The different representations \refeq{eq:philin}
and \refeq{eq:phinl} are physically equivalent \cite{stue2,stue3}, i.e.\ both
yield the same S-matrix. Inserting \refeq{eq:phinl} into the Lagrangian
\refeq{eq:smlag}, one obtains
\beqar
\lag &=& - \frac{1}{2}\tr{W_{\mu\nu}W^{\mu\nu}}
- \frac{1}{4}B_{\mu\nu}B^{\mu\nu}
+ \frac{1}{4}(v+H)^2\tr{(D_\mu U)^\dagger(D^\mu U)} \nn\\
&&{} + \frac{1}{2}(\partial_\mu H)(\partial^\mu H)
+ \frac{1}{2}\mu^2 (v+H)^2
-\frac{1}{16}\lambda(v+H)^4.
\label{eq:smlag2}
\eeqar
In this form the advantage of the non-linear representation of $\Phi$ is
apparent. Owing to the unitarity of $U$ the unphysical Goldstone field
$\vp$ only enters the kinetic term of the scalar fields, but drops out
in the cubic and quartic
scalar self
interactions.

Our conventions and notation for the parameters
and fields follow the ones of \citeres{bfm4,bfm5,de93}. Moreover,
substituting $g_2\to g$, $g_1\to 0$, $B^\mu\to 0$
reproduces the results of \citere{sdcgk} for the pure SU(2) theory.

Finally, we consider the case of a very heavy Higgs boson, i.e.\ the
limit $\MH\to\infty$. At tree level,
the Lagrangian \refeq{eq:smlag2} reduces to
the one of the {\it gauged non-linear $\si$-model}
(GNLSM) \cite{apbe,long}, which follows from
\refeq{eq:smlag2} simply by disregarding the field $H$. Beyond tree
level the situation is much more complicated, as loop corrections
associated with virtual Higgs-boson exchange lead to additional
(effective) interactions. Our aim is to integrate out the heavy Higgs
field at one loop and to construct the
corresponding one-loop effective Lagrangian. However, the Lagrangian
\refeq{eq:smlag2} contains the field $H$ up to quartic power so that
Gaussian integration is not directly applicable in the path integral. At
one loop this problem is circumvented by the
background-field method (BFM).

\subsection{The background-field method}
\label{ssec:bfm}

The BFM
\cite{bfm1,bfm2}
was applied to the SM with linearly
realized Higgs sector in
\citeres{bfm3,bfm4,bfm5}.
For a pure SU(2) gauge
theory we generalized the BFM to the non-linear representation of the
scalar sector in \citere{sdcgk}. The same procedure also applies to the
$\mathrm{SU(2)_{\mathrm W}
\times U(1)_{\mathrm Y}}$ SM. Accordingly, we split the fields
into background and quantum fields as follows:
\beq
W^\mu \to \hat{W}^\mu+W^\mu, \quad
B^\mu \to \hat{B}^\mu+B^\mu, \quad
H \to \hat{H}+H, \quad U \to \hat{U}U,
\label{eq:split}
\eeq
where the hats mark background fields. In opposite to the gauge and
Higgs fields the matrix $U$ \refeq{eq:phinl},
which contains the Goldstone field $\vp$, is split multiplicatively.
Recall that only the quantum fields are
quantized, i.e.\ they represent variables of integration in the path
integral. The background fields act as sources for the generation of
vertex functions in the effective action.
The background fields correspond to tree lines and the quantum fields
to lines in loops. Thus, at one loop only the part of the Lagrangian
quadratic in the quantum fields is relevant, and therefore Gaussian
integration is applicable. Furthermore, this means that
for the construction of
vertex functions only the gauge of the quantum fields has to be fixed.
Choosing the gauge-fixing term for the quantum fields such that gauge
invariance with respect to the background fields is retained, the
effective action is ``background-gauge-invariant'', too. For the linearly
realized Higgs sector \refeq{eq:philin} an appropriate gauge-fixing term
was given in \citeres{bfm3,bfm4,bfm5}, for the non-linear case
\refeq{eq:phinl} we use
\beq
\lag_{\mathrm{gf}}=
-\frac{1}{\xi_W} \tr{\left(\DW^\mu W_\mu
+\frac{1}{2}\xi_W g_2v\hat{U}\vp \hat{U}^\dagger\right)^2}
-\frac{1}{2\xi_B}\left(\partial^\mu B_\mu+\frac{1}{2}\xi_B g_1v
\vp_3
\right)^2
\label{eq:gfterm}
\eeq
with
\beq
\DW^\mu X = \partial^\mu X-ig_2[\hW^\mu,X],
\label{eq:DW}
\eeq
which is the natural extension of the choice made in \citere{sdcgk} for
the SU(2) model. In the following we set $\xi=\xi_W=\xi_B$ in order to
avoid mixing between the neutral gauge fields $A$, $Z$ at tree level. It
is straightforward to check that Lagrangian \refeq{eq:smlag2} with
$\lag_{\mathrm{gf}}$ of \refeq{eq:gfterm} leads to an effective action which
is invariant under the following background gauge transformation:
\beq
\hat{W}^\mu \to
S\left(\hat{W}^\mu+\frac{i}{g_2}\partial^\mu\right)S^\dagger, \quad
\hat{B}^\mu \to \hat{B}^\mu+\partial^\mu\theta_{\mathrm Y},  \quad
\hat{H} \to \hat{H}, \quad
\hat{U} \to S\hat{U}\SY \quad
\label{eq:bgtrafo}
\eeq
with
\beq
S=\exp\left(ig_2\theta\right), \qquad
\SY=\exp\left(ig_1\theta_{\mathrm Y}\frac{\tau_3}{2}\right),
\label{eq:S}
\eeq
associated with the following substitution of the quantum fields in the
path integral:
\beq
W^\mu \to S W^\mu S^\dagger, \quad
B^\mu \to B^\mu,  \quad
H \to H, \quad
U \to \SY^\dagger U\SY. \quad
\eeq
$\theta=\theta_i\tau_i/2$ and $\theta_{\mathrm Y}$ denote the group
parameters of the $\mathrm{SU(2)_{\mathrm W}}$ and $\mathrm{U(1)_{\mathrm
Y}}$,
respectively.

The Faddeev--Popov Lagrangian $\lag_{\mathrm{ghost}}$, which corresponds to
the gauge-fixing term \refeq{eq:gfterm}, is constructed as usual.
In particular, $\lag_{\mathrm{ghost}}$ neither involves the quantum nor the
background Higgs field.

\subsection{The Stueckelberg formalism}
\label{ssec:stf}

The gauge of the background fields has not been specified so far and can
be chosen independently from the one of the quantum fields. It is most
convenient to choose the {\it unitary gauge} (U-gauge) for the
background fields, where all background Goldstone fields disappear. To
this end, we use the Stueckelberg formalism
\cite{stue1,stue2,stue4,stue3}, which has been generalized to the BFM
in \citeres{sdcgk,chey}. We apply the Stueckelberg transformation
\beq
\hat{W}^\mu \to \hat{U}\hat{W}^\mu\hat{U}^\dagger +\frac{i}{g_2}
\hat{U}\partial^\mu\hat{U}^\dagger, \qquad
\hat{B}^\mu \to \hat{B}^\mu,\qquad
W^\mu \to \hat{U}W^\mu\hat{U}^\dagger,
\qquad B^\mu \to B^\mu,
\label{eq:sttrafo}
\eeq
which transforms the $W$
field-strength and covariant derivative as
\beq
D^\mu\hat{U}U \to \hat{U} D^\mu U,\qquad
(\hat{W}^{\mu\nu}+W^{\mu\nu})\to \hat{U}
(\hat{W}^{\mu\nu}+W^{\mu\nu}) \hat{U}^\dagger.
\label{sttrafo2}
\eeq
The effect of this transformation on the Lagrangian is to map the matrix
$\hat{U}$ to the unit matrix ($\hat{U}\to{\bf 1}$), but leaving
everything else unaffected. The fact that no background Goldstone fields
are present in intermediate steps of the heavy-Higgs expansion
simplifies our calculation drastically. Inverting the Stueckelberg
transformation \refeq{eq:sttrafo} at the end,
we recover
the result for an arbitrary background gauge.

\section{Diagonalizing the Higgs part of the one-loop Lagrangian}
\label{sec:diag}

As pointed out above,
at one loop only those terms of the Lagrangian are relevant which are
bilinear in the quantum fields. In the background U-gauge the full
one-loop Lagrangian reads
\beqar
\lag^{\mathrm{1-loop}} &=&
\tr{ W_\mu\left( g^{\mu\nu} \DW^2
+\frac{1-\xi}{\xi}\DW^\mu\DW^\nu +2ig_2 \hat{W}^{\mu\nu}\right) W_\nu} \nn\\
&&{}+\frac{1}{2} B_\mu\left( g^{\mu\nu} \partial^2
+\frac{1-\xi}{\xi}\partial^\mu\partial^\nu \right) B_\nu
+\frac{1}{4}g_2^2(v+\hat{H})^2 \tr{C_\mu C^\mu}\nn\\
&&{}-\tr{\vp\left(\frac{1}{v^2}\hat{D}_{\mu}(v+\hH)^2\hat{D}^\mu
+\frac{1}{4}\xi{g_2^2v^2}
+g_2^2\frac{1}{v^2}(v+\hH)^2
\hC_\mu \hC^\mu\right)\vp}
-\frac{1}{8}\xi g_1^2 v^2 \vp_3^2\nn\\
&&{}-\frac{1}{2}H\left(\partial^2-\mu^2
+\frac{3}{4}\la(v+\hH)^2
-\frac{1}{2}g_2^2\tr{\hC_\mu \hC^\mu}\right)H\nn\\
&&{}-2 g_2\frac{1}{v}(v+\hH) H \tr{\hC_\mu\partial^\mu\vp}
-2ig_1g_2\frac{1}{v}(v+\hH)H \tr{\vp\hW_\mu\tau_3}\hB^\mu
\nn\\
&&{}+g_2^2 (v+\hH) H \tr{\hC_\mu C^\mu}
-g_2 \frac{1}{v} (2v+\hH) \hH \tr{C_\mu \partial^\mu\vp}
\nn\\
&&{}-2ig_2^2v\tr{W_\mu\hW^\mu\vp}+ig_1 g_2\frac{1}{v}
(v+\hH)^2 \left(B_\mu\tr{\tau_3\hW^\mu\vp}
+\hB_\mu\tr{\tau_3 W^\mu\vp}\right)
\nn\\
&&{}+ \lag_{\mathrm{ghost}}.
\label{eq:lag}
\eeqar
The auxiliary background field $\hC^\mu$ occurring in
\refeq{eq:lag} is defined via
\beq
\hC^\mu \,=\, \hW^\mu+\frac{g_1}{g_2}\hB^\mu\frac{\tau_3}{2}
\,=\, \frac{1}{2}\left(\hW^\mu_1{\tau_1} + \hW^\mu_2{\tau_2}
+\frac{1}{\cw}\hZ^\mu{\tau_3}\right)
\eeq
and the corresponding quantum field
analogously.

Since the ghost Lagrangian $\lag_{\mathrm{ghost}}$ is bilinear in the
Faddeev-Popov ghost
fields, which do not have a background part,
the one-loop part of $\lag_{\mathrm{ghost}}$
in \refeq{eq:lag} contains no other
quantum fields than ghosts and remains unaffected by all following
manipulations.

Fortunately, not all terms of $\lag^{\mathrm{1-loop}}$ in \refeq{eq:lag}
are relevant for the
construction of the effective Lagrangian describing the non-decoupling
effects. In the following we only consider contributions of
$\OH{0}$, i.e.\ we neglect all terms which
yield no effects in the limit
$\MH\to\infty$. Our complete method for the $1/\MH$-expansion was
described in detail in \citere{sdcgk} for the SU(2) case. Thus, here we
shorten the presentation to the most important steps and omit
more technical details. We write the one-loop Lagrangian in the
symbolic form
\beqar
\lag^{\mathrm{1-loop}}&=&
-\frac{1}{2}H\,\Delta_{\ss H}\,H
+H\,\tr{X^\mu_{\ss H\OW}\,\OW_\mu}
+H\,\tr{X_{\ss H\vp}\,\vp}
\nn\\ && {}
+\tr{\OW_\mu\,\Delta^{\mu\nu}_{\ss \OW}\,\OW_\nu}
+\frac{1}{2}\tr{A_\mu\,\Delta^{\mu\nu}_{\ss A}\,A_\nu}
+\tr{A_\mu\,X^{\mu\nu}_{\ss A\OW}\,\OW_\nu}
\nn\\ && {}
-\tr{\vp\,\Delta_{\ss \vp}\,\vp}
+\tr{\OW_\mu\,X^\mu_{\ss \OW\vp}\,\vp}
+\tr{A_\mu\,X^\mu_{\ss A\vp}\,\vp}
\quad+\lag_{\mathrm{ghost}}
\label{eq:lag2}
\eeqar
with the modified quantum
SU(2)$_\PW$ field
\beq
\OW^\mu=\frac{1}{2}\left(W^\mu_1\tau_1+W^\mu_2\tau_2+Z^\mu\tau_3\right)
\eeq
and the quantum photon field $A^\mu$. Obviously, there is no $AH$-term
in \refeq{eq:lag}.

Applying Gaussian integration over $H$ in the path integral directly
to $\lag^{\mathrm{1-loop}}$ of \refeq{eq:lag2},
the terms linear in the quantum Higgs field $H$ would yield (problematic)
terms with inverse operators acting on quantum fields. However,
the terms linear in $H$ can be removed by appropriate shifts
of the quantum fields \cite{sdcgk,gale,chey}. Substituting successively
\cite{sdcgk}
\beqar
\vp &\;\to\;& \vp
+\frac{1}{2}\hat\Delta_\vp^{-1}X_{\ss H\vp}^\dagger\,H
+\frac{1}{2}\hat\Delta_\vp^{-1}X_{\ss \OW\vp}^{\mu\dagger}\,\OW_\mu,
\nn\\
\OW^\mu &\;\to\;& \OW^\mu
-\frac{1}{2} \hat{\tilde{\Delta}}^{-1\mu\nu}_{\ss\OW}
\tilde{X}_{{\ss H\OW},\nu}^\dagger H,
\nn\\
\vp &\;\to\;& \vp
-\frac{1}{2}\hat\Delta_\vp^{-1}X_{\ss \OW\vp}^{\mu\dagger}\,\OW_\mu
\label{eq:shifts}
\eeqar
with
\beqar
\tilde{\Delta}^{\mu\nu}_{\OW} &=& {\Delta}^{\mu\nu}_{\OW} +\frac{1}{4}
X^\mu_{\ss \OW\vp}\hat\Delta_\vp^{-1}
X^{\nu\dagger}_{\ss \OW\vp},\nn\\
{\tilde{X}}_{\ss H \OW,\mu}&=&
{X}_{{\ss H\OW},\mu}+\frac{1}{2}X_{\ss
H\vp}\hat\Delta^{-1}_{\vp}X_{{\ss\OW\vp},\mu}^\dagger
\label{eq:deftilde}
\eeqar
completely eliminates the $H\OW$- and $H\vp$-terms without
changing the $\OW\vp$-mixing. The bilinear $H$-operator transforms into
\beq
\Delta_{\ss H} \;\to\;
\tilde{\Delta}_{\ss H} =
\Delta_{\ss H}
-\frac{1}{2}\tr{X_{\ss H\vp}\hat\Delta^{-1}_{\vp}X^\dagger_{\ss H\vp}}
+\frac{1}{2}\tr{
\tilde{X}_{{\ss H\OW},\mu}
\hat{\tilde{\Delta}}^{-1\mu\nu}_{\OW}
\tilde{X}_{{\ss H\OW},\nu}^\dagger}.
\eeq
The meaning of the hats over the inverse operators
will be explained below.
In contrast to the SU(2) case, the transformations \refeq{eq:shifts}
produce mixing terms between the quantum Higgs field $H$ and the
photon field $A$. Analogously to \refeq{eq:shifts}, these $AH$-terms
can also be removed by suitable (but more involved) shifts without
affecting the $H$-independent contributions. Only
${\tilde{\Delta}}_{\ss H}$ is modified again. However, these
additional terms in ${\tilde{\Delta}}_{\ss H}$
only yield $\OH{-2}$-contributions in the subsequent $1/\MH$-expansion,
and thus are not
explicitly
discussed here. This can easily be seen as follows:
In \citere{sdcgk} it has been shown that the Yang-Mills couplings
and the vector-Goldstone term yield no
$\OH{0}$-contributions
when integrating out the Higgs field
and can thus be
neglected. However, the quantum photon field $A$ only couples to the
other quantum fields through the Yang-Mills and the
vector-Goldstone term. Thus, at $\OH{0}$ this field may be dropped in
\refeq{eq:lag2} from the beginning.
At the diagrammatical level this means that there are no
$\OH{0}$-contributions from loops with both photon and Higgs
fields, which is in
accordance with the diagrammatical calculation in \citere{hemo}.
Taking only into account effects
of $\OH{0}$, $\tilde{\Delta}_{\ss H}$ reduces to
\beq
\tilde{\Delta}_{\ss H}\to
\tilde{\tilde{\Delta}}_{\ss H}=
\Delta_{\ss H}
-\frac{1}{2}\tr{X_{\ss H\vp}\hat\Delta^{-1}_{\vp}X^\dagger_{\ss H\vp}}
+\frac{1}{2}\tr{X_{{\ss H\OW},\mu}
\hat{{\Delta}}^{-1\mu\nu}_{\ss\OW,0}
X_{{\ss H\OW},\nu}^\dagger}
\label{eq:ttDH}
\eeq
as in \citere{sdcgk}.
In \refeq{eq:ttDH} we already made use of the fact that only the
lowest-order part $\Delta_{\ss\OW,0}^{\mu\nu}$ of
$\Delta_{\ss\OW}^{\mu\nu}$ contributes in $\OH{0}$,
in analogy to the situation
in the SU(2) case.

We still have to supply the meaning of the hat over the inverse
operators in the previous formulas. As in \citere{sdcgk},
$\hat{\Delta}^{-1}$ denotes the restriction of the hermitian,
$2\times 2$-matrix-valued inverse operator $\Delta^{-1}$ to the subspace
spanned by the Pauli matrices $\tau_i$.
Only with this restriction the shifts \refeq{eq:shifts} make sense,
because it ensures that the rhs of these shifts  are linear
combinations of the Pauli matrices \cite{sdcgk}.
In terms of a perturbative expansion $\hat{\Delta}^{-1}$ is given by
\beqar
\hat\Delta^{-1} &=& \disp
\Delta_0^{-1}P\;\sum_{n=0}^{\infty}\;(-\Pi\Delta_0^{-1}P)^n \nn\\
&=& \Delta_0^{-1}P \;-\; \Delta_0^{-1}P\Pi\Delta_0^{-1}P \;+\;
\Delta_0^{-1}P\Pi\Delta_0^{-1}P\Pi\Delta_0^{-1}P \;-\; \cdots,
\label{eq:invDel}
\eeqar
where $\Delta_0$ denotes the lowest-order contribution (which is
proportional to the unit matrix) to the full operator
$\Delta=\Delta_0+\Pi$. The operator $P$ is
the projector onto the subspace spanned by
the $\tau_i$. More generally, we define
\beqar
P_i X &=& \frac{1}{2} \tau_i \tr{\tau_i X}
\qquad\mbox{(no summation over $i$),} \nn\\
P &=& \sum_{i=1}^{3} P_i,
\label{eq:P}
\eeqar
where the $P_i$ project on the single Pauli matrices $\tau_i$,
respectively.

For the operators $\Delta$, $X$ of the one-loop Lagrangian \refeq{eq:lag2}
we just give the terms which are relevant for
${\tilde{\tilde{\Delta}}}_{\ss H}$ in \refeq{eq:ttDH}, namely
\beqar
\Delta_{\ss \OW,0}^{\mu\nu} &=&
g^{\mu\nu}\partial^2+\frac{1-\xi}{\xi}\partial^\mu\partial^\nu
+g^{\mu\nu}\MW^2\left(1+\frac{\sw^2}{\cw^2}P_3\right),
\nn\\
\Delta_{\ss \vp} &=&
\hD^\mu\left(1+\frac{\hH}{v}\right)^2\hD_\mu
+g_2^2\hC^\mu\hC_\mu\left(1+\frac{\hH}{v}\right)^2
+\xi\MW^2\left(1+\frac{\sw^2}{\cw^2}P_3\right),
\nn\\
\Delta_{\ss H} &=&
\partial^2+\MH^2
+\frac{3}{2}\MH^2\frac{\hH}{v}\left(2+\frac{\hH}{v}\right)
-\frac{1}{2}g_2^2 \tr{\hC^\mu\hC_\mu},
\nn\\
X_{\ss H \OW}^\mu &=&
2g_2\left(1+\frac{\hH}{v}\right)
\MW\hC^\mu\left(1+\frac{1-\cw}{\cw}P_3\right),
\nn\\
X_{\ss H \vp} &=&
2g_2\left(1+\frac{\hH}{v}\right)
\left(-\hC^\mu\partial_\mu+ig_1\hB^\mu\tau_3\hW_\mu\right).
\label{eq:DelX}
\eeqar

After all these manipulations the resulting one-loop Lagrangian is
obtained from \refeq{eq:lag2} upon disregarding $X^\mu_{\ss H\OW}$,
$X_{\ss H\vp}$ and replacing $\Delta_{\ss H}$ by
$\tilde{{\tilde{\Delta}}}_{\ss H}$ of \refeq{eq:ttDH}, where
terms yielding only $\OH{-2}$-contributions are neglected.

\section{Integrating out the quantum Higgs field and
{\protect\boldmath${1}/{\MH}$}-expansion}
\label{sec:helim}

The next step is to perform the path integral over
the quantum field $H$ by Gaussian integration.
For a detailed discussen of this procedure, we
again refer to \citere{sdcgk}. The term quadratic in $H$ yields
a functional determinant which can be expressed
in terms of an effective Lagrangian \cite{sdcgk,chan}
\beq
\leff = \frac{i}{2}
\intdp \log \left(\tilde{\tilde{\Delta}}_H\dpp\right).
\label{eq:leff}
\eeq
$\tilde{\tilde{\Delta}}_H\dpp$ can be expanded in terms of
derivatives%
\footnote{The first line of \refeq{eq:Dexpand} cannot be taken
literally for the derivative expansion. The partial derivatives do
not commute with the background fields in
$\tilde{\tilde{\Delta}}_H(x,ip)$, and thus one also has to take care
of the position of the derivative operators, which can easily be
achieved in the actual calculation.},
\beqar
\tilde{\tilde{\Delta}}_H\dpp &=&
\sum_{n=0}^\infty \frac{(-i)^n}{n!}
\left[\frac{\partial^n}
{\displaystyle\partial{p_{\mu_1}}\ldots\partial{p_{\mu_n}}}
\tilde{\tilde{\Delta}}_H(x,ip)\right]
\partial_{\mu_1}\ldots\partial_{\mu_n}
\nn\\[.3em]
&=& -p^2+\MH^2+\Pi(x,p,\partial_x),
\label{eq:Dexpand}
\eeqar
leading to the following expansion of the logarithm
\beq
\log \tilde{\tilde{\Delta}}_H\dpp
= \log(-p^2+\MH^2)
-\sum_{n=1}^\infty \frac{1}{n}\left(\frac{\Pi}{p^2-\MH^2}\right)^n.
\label{eq:logexpand}
\eeq
The first log-term of \refeq{eq:logexpand} yields a constant contribution to
the effective Lagrangian, which is irrelevant in this context and will
be dropped in the following. The powers of $\Pi$ in \refeq{eq:logexpand}
contain propagator terms $(p^2-M^2)^{-m}$
with $M^2=\MW^2,\MZ^2,\xi\MW^2$ or $\xi\MZ^2$
originating from the derivative
expansion of the inverse propagators $\hat\Delta^{-1}_{\vp}$,
$\hat{{\Delta}}^{-1\mu\nu}_{\ss\OW,0}$. Hence, upon inserting
expansion \refeq{eq:logexpand} into \refeq{eq:leff}, the effective
Lagrangian can be expressed in terms of one-loop vacuum integrals of the
type
\beq
I^i_{klm}(\xi) \, g_{\mu_1\ldots\mu_{2k}}
= \frac{(2\pi\mu)^{4-D}}{i\pi^2}
\int {d^D p} \frac{\disp
p_{\mu_1}\ldots p_{\mu_{2k}}}{(p^2-\MH^2)^l(p^2-\xi M_i^2)^m},
\qquad M_i=\MW,\MZ.
\label{eq:intnot}
\eeq
In \refeq{eq:intnot} it is already indicated that we use dimensional
regularization throughout with $D$ denoting the number of space-time
dimensions, and $\mu$ representing the reference mass scale.
$g_{\mu_1\ldots\mu_{2k}}$ is the totally symmetric tensor
of rank $2k$ built of the metric tensor $g_{\mu\nu}$.
For $D\to 4$ the integrals $I^i_{klm}(\xi)$ are $\O(\MH^n)$ with
\beq
n=4+2(k-l-m)
\label{eq:cond}
\eeq
if $n\ge0$, and $\OH{-2}$ or less if $n<0$.
The explicit expressions for the integrals relevant
for $\leff$ are listed in \refapp{app:ints}.
In particular, the
$\OH{0}$-parts of all logarithmically divergent
integrals
are independent of $\xi$ and $M_i^2$. Consequently, the index
$i$ and the argument $\xi$ will be dropped for these in the following.
In addition to the $\MH$-dependence of the integrals,
there is an explicit $\MH$-dependence in the generated
effective Lagrangian due to the Higgs self interactions and an implicit
$\MH$-dependence stemming from the occurrence
of the background Higgs field $\hH$ which will later be eliminated
by a propagator expansion yielding $\hH=\OH{-2}$. Thus, as in
\citere{sdcgk}, we introduce an auxiliary power-counting parameter
$\c$, which counts the powers of $p_\mu$, $\hH$ and $\MH$
according to
\beq
p_\mu \to \c,\qquad \MH \to \c, \qquad \hH \to \c^{-2}.
\label{eq:count}
\eeq
In order to obtain the effective Lagrangian
at $\OH{0}$,
we only have to consider contributions up to $\O(\c^{-4})$
in the expansion of $\log \tilde{\tilde{\Delta}}_H\dpp$ (i.e.\ up to
$\ord{-2}$ in $\tilde{\tilde{\Delta}}_H\dpp$) and can neglect higher
negative powers of $\c$.

As a result of this power counting it turns out that most of the
contributions of the projection operator $P_3$ \refeq{eq:P} in
$\Delta_{\ss\OW,0}^{\mu\nu}$ and $\Delta_{\ss\vp}$
\refeq{eq:DelX} can be neglected
at $\OH{0}$. In order to illustrate this, we consider the
operator $\hat{\Delta}_{\ss\vp}^{-1}\dpp$,
which occurs in \refeq{eq:leff} with \refeq{eq:ttDH}.
Using \refeq{eq:P} we write
\beq
\MW^2\left(1+\frac{\sw^2}{\cw^2}P_3\right)
P
=M_i^2 P_i, \qquad
\mbox{with} \qquad \quad M_{1,2}=\MW, \quad M_3=M_Z
\label{eq:msps}
\eeq
and find with \refeq{eq:invDel}
\beqar
\hat{\Delta}_{\ss\vp}^{-1}\dpp&=&{}-\frac{1}{p^2-\xi M_{i}^2}P_i\nn\\
&&{}-\frac{1}{(p^2-\xi M_{i}^2)(p^2-\xi M_{j}^2)}
P_{i}\left[-\frac{2\hH}{v}p^2
+2ip_\mu\hD^\mu+\hD^2+g_2^2 \hC^\mu \hC_\mu\right]
P_{j}\nn\\
&&{}+\frac{1}{(p^2-\xi M_{i}^2)(p^2-\xi M_{j}^2)
(p^2-\xi M_{k}^2)}
4 P_{i}
(p_\mu\hD^\mu)P_{j}
(p_\nu\hD^\nu)P_{k}\nn\\
&&{}+\ord{-5}.
\label{eq:dpinv}
\eeqar
The operator $(p^2-\xi M_{i}^2)^{-1}P_i$ occurring several
times in this expression can be written as
\beq
\frac{1}{p^2-\xi M_{i}^2}P_i
=\frac{1}{p^2-\xi \MW^2} P - \xi\frac{\MW^2-\MZ^2}{(p^2-\xi
\MW^2)(p^2-\xi\MZ^2)}P_3.
\label{eq:P3}
\eeq
The second term in \refeq{eq:P3} is $\ord{-4}$ and can thus be
neglected in the second and the third term of \refeq{eq:dpinv},
because $\hat{\Delta}_{\ss\vp}^{-1}\dpp$ is only needed at $\ord{-4}$.

Expanding $\log\tth$ and integrating over  $p$
in analogy to \citere{sdcgk}, we find the effective Lagrangian
according to \refeq{eq:leff}, \refeq{eq:Dexpand} and \refeq{eq:logexpand}:
\beqar
\leff&=\displaystyle\fac \Bigg\{&\phantom{+}\,\,\,I_{010}
\left[\frac{3g_2\MH^2}{4\MW}\hH
+\frac{3g_2^2\MH^2}{16 \MW^2}\hH^2-\frac{1}{4}
g_2^2\tr{\hC_\mu \hC^\mu}\right]\nn\\
&&{}- I_{011}\;g_2^2M_i^2\tr{\hC_\mu P_i \hC^\mu}\nn\\
&&{}+ I_{111}^i(1)\;g_2^2\tr{\hC_\mu P_i \hC^\mu}\nn\\
&&{}+ I_{011}g_2^2\left[\tr{(\partial_\mu \hC^\mu)^2}
+2ig_1\hB_\mu\tr{\tau_3\hW^\mu(\partial_\nu\hC^\nu)}
\right.\nn\\
&&\qquad\qquad\qquad\qquad\qquad\left.{}
+g_1^2\hB_\mu \hB_\nu\tr{\tau_3\hW^\mu P
\hW^\nu\tau_3}\right]\nn\\
&&{}+ I_{112}g_2^2\left[-4\tr{(\partial_\mu\hC^\mu)(\hD_\nu\hC^\nu)}
+\tr{\hC_\mu \hD^2 \hC^\mu} \right.\nn\\
&&\qquad\qquad\qquad\qquad\qquad\left.
{}+g_2^2 \tr{\hC_\mu \hC^\mu \hC_\nu
\hC^\nu}-4ig_1\hB_\mu\tr{\tau_3\hW^\mu P \hD_\nu\hC^\nu}
\right]\nn\\
&&{}- I_{213}4g_2^2\left[\tr{\hC_\mu \hD^\mu P \hD^\nu
\hC_\nu}+\tr{\hC_\mu \hD^\nu P \hD^\mu
\hC_\nu}
+\tr{\hC_\mu \hD_\nu P \hD^\nu
\hC^\mu}\right]\nn\\
&&{}+ I_{020}\left[\frac{9g_2^2\MH^4}{16\MW^2}\hH^2
-\frac{3g_2^3\MH^2}{8\MW}\hH \tr{\hC_\mu \hC^\mu} +
\frac{1}{16} g_2^4\left(\tr{\hC_\mu \hC^\mu}\right)^2\right]\nn\\
&&{}+ I_{121}\left[\frac{3g_2^3\MH^2}{2\MW}\hH \tr{\hC_\mu
\hC^\mu}-\frac{1}{2}g_2^4 \left(\tr{\hC_\mu
\hC^\mu}\right)^2\right]\nn\\
&&{}+ I_{222}g_2^4\left[\left(\tr{\hC_\mu
\hC^\mu}\right)^2 + 2\left(\tr{\hC_\mu \hC_\nu}\right)^2
\right]\Bigg\}\nn\\
&&{}\back+\ord{-2},
\label{eq:leff1}
\eeqar
where we have used the notation \refeq{eq:intnot} for the
(vacuum) one-loop integrals.

The origin of the various terms in \refeq{eq:leff1} is the following:
The first line is the contribution of $\Delta_{\ss H}$ in \refeq{eq:ttDH},
the second stems from of $X_{{\ss H\OW},\mu}
\hat{{\Delta}}^{-1\mu\nu}_{\ss\OW,0}
X_{{\ss H\OW},\nu}^\dagger$, the third gets contributions from
$X_{\ss H\vp} \hat\Delta^{-1}_{\vp}X^\dagger_{\ss H\vp}$ and
$X_{{\ss H\OW},\mu} \hat{{\Delta}}^{-1\mu\nu}_{\ss\OW,0}
X_{{\ss H\OW},\nu}^\dagger$ together, and the remaining terms come from
$X_{\ss H\vp} \hat\Delta^{-1}_{\vp}X^\dagger_{\ss H\vp}$.

\section{Introducing standard traces
and inverting the Stueckelberg transformation}
\label{sec:invstue}

The effective Lagrangian \refeq{eq:leff1} has to be
written in a more convenient form.
Since we want to invert the Stueckelberg transformation
\refeq{eq:sttrafo} in order to obtain $\leff$ in an arbitrary
background gauge, it is useful to introduce appropriate
gauge-invariant standard traces. Such traces have for instance been
introduced in \citere{long}\footnote{In \citere{long}
the couplings constants $\alpha_i$ are part of the effective terms
$\lag_i$ while here they are not. Apart from this, our terms are
identical with those used in \citere{long}. The $\lag^\prime_1$
defined there
corresponds to our $\lag_0$, and the traces in $\lag_6$, \dots,
$\lag_{10}$, $\lag_{12}$ and $\lag_{13}$ of \citere{long} do not
occur in our calculation and thus are not listed here.}.
Since we presently work in the U-gauge for the background fields,
we express these terms both in their gauge-invariant form
(lhs of the arrow) and in the U-gauge (rhs of the arrow):
\beqar
\lag_0&\,=\,&\rlap{${\MW^2}\left(\tr{\hT \hV_\mu}\right)^2$}
\phantom{\frac{1}{2}i g_2\hB_{\mu\nu}\tr{\hT[\hV^\mu,\hV^\nu]}\,}
\toug\, -{g_2^2\MW^2}\left(\tr{\tau_3 \hC_\mu}\right)^2
,\nn\\
\lag_1&=&\rlap{$\displaystyle
\frac{1}{2}g_2^2 \hB_{\mu\nu}\tr{\hT\hW^{\mu\nu}}$}
\phantom{\frac{1}{2}i g_2\hB_{\mu\nu}\tr{\hT[\hV^\mu,\hV^\nu]}\,}
\toug\,
\frac{1}{2}g_2^2 \hB_{\mu\nu}\tr{\tau_3 \hW^{\mu\nu}}
,\nn\\
\lag_2&=&
\rlap{$\displaystyle\frac{1}{2}i
  g_2\hB_{\mu\nu}\tr{\hT[\hV^\mu,\hV^\nu]}$}
\phantom{\frac{1}{2}i g_2\hB_{\mu\nu}\tr{\hT[\hV^\mu,\hV^\nu]}\,}
\toug\,
-\frac{1}{2}i
  g_2^3\hB_{\mu\nu}\tr{\tau_3[\hC^\mu,\hC^\nu]}
,\nn\\
\lag_3&=&\rlap{$ig_2\tr{\hW_{\mu\nu}[\hV^\mu,\hV^\nu]}$}
\phantom{\frac{1}{2}i g_2\hB_{\mu\nu}\tr{\hT[\hV^\mu,\hV^\nu]}\,}
\toug \,
-ig_2^3\tr{\hW_{\mu\nu}[\hC^\mu,\hC^\nu]}
,\nn\\
\lag_4&=&\rlap{$\left(\tr{\hV_\mu\hV_\nu}\right)^2$}
\phantom{\frac{1}{2}i g_2\hB_{\mu\nu}\tr{\hT[\hV^\mu,\hV^\nu]}\,}
\toug\,
g_2^4\left(\tr{\hC_\mu\hC_\nu}\right)^2
,\nn\\
\lag_5&=&\rlap{$\left(\tr{\hV_\mu\hV^\mu}\right)^2$}
\phantom{\frac{1}{2}i g_2\hB_{\mu\nu}\tr{\hT[\hV^\mu,\hV^\nu]}\,}
\toug \,
g_2^4\left(\tr{\hC_\mu\hC^\mu}\right)^2
,\nn\\
\lag_{11}&=&\rlap{$\tr{\left(\hD_W^\mu\hV_\mu\right)^2}$}
\phantom{\frac{1}{2}i g_2\hB_{\mu\nu}\tr{\hT[\hV^\mu,\hV^\nu]}\,}
\toug \,
-g_2^2\tr{\left(\hD_W^\mu\hC_\mu\right)^2}
\label{eq:sttr}
\eeqar
with $\hD_W$ defined in \refeq{eq:DW}. Following \citere{long}, we
introduce the shorthand notation
\beq
\hV^\mu=\left(\hD^\mu\hU\right)\hU^\dagger, \qquad \hT=\hU \tau_3
\hU^\dagger.
\label{eq:V}
\eeq

First, we
consider the terms in \refeq{eq:leff1} which contain
derivatives or covariant derivatives \refeq{eq:dphi}.
These terms are proportional to $I_{011}$, $I_{112}$ or $I_{213}$.
We express the derivatives in terms of
field-strength tensors \refeq{eq:fst}
and vector-covariant derivatives $\hD_W^\mu$ \refeq{eq:DW}.
These terms become
\beqar
\leff\bigg|_{I_{011}}^{\mathrm{deriv}}&=&-\fac I_{011}\;\lag_{11},\nn\\
\leff\bigg|_{I_{112}}^{\mathrm{deriv}}&=&\fac I_{112}
\bigg[-\frac{1}{2}
g_2^2\tr{\hW_{\mu\nu}\hW^{\mu\nu}}-\frac{1}{4}g_1^2\hB_{\mu\nu}
\hB^{\mu\nu}\nn\\
&&\qquad\qquad\qquad\qquad
{}-\frac{g_1}{g_2}\lag_1-\frac{1}{2}\frac{g_1}{g_2}\lag_2+
\frac{1}{2}\lag_3
-\frac{1}{2}\lag_5
+5\lag_{11}\bigg],\nn\\
\leff\bigg|_{I_{213}}^{\mathrm{deriv}}&=&\fac I_{213}
\bigg[2g_2^2\tr{\hW_{\mu\nu}\hW^{\mu\nu}}+g_1^2\hB_{\mu\nu}
\hB^{\mu\nu}\nn\\
&&\qquad\qquad\qquad\qquad
{}+4\frac{g_1}{g_2}\lag_1+4\frac{g_1}{g_2}\lag_2-4\lag_3-4\lag_4+4
\lag_5-12\lag_{11}\bigg].
\label{eq:derterms}
\eeqar
Next, we consider the terms proportional to $I_{011}$ and
$I^i_{111}(1)$ which contain the operators $P_i$ \refeq{eq:P}
with different coefficients for $i=1,2$ and $i=3$.
These can easily be evaluated by using
\beq
M_i^2\tr{\hC_\mu P_i \hC^\mu}
=\MW^2\tr{\hC_\mu \hC^\mu}
+\frac{1}{2}\frac{g_1^2}{g_2^2}\MW^2\left(\tr{\tau_3\hC_\mu}\right)^2
\eeq
and a corresponding identity for $I^i_{111}(1) \tr{\hC_\mu P_i
\hC^\mu}$. We find:
\beqar
\leff\bigg|^{P_i}_{I_{011}}&=&\fac I_{011}\left[-g_2^2\MW^2
\tr{\hC_\mu\hC^\mu}+\frac{1}{2}\frac{g_1^2}{g_2^2}\lag_0\right],\nn\\
\leff\bigg|^{P_i}_{I^i_{111}(1)}&=&\fac \left[ I^W_{111}(1)\;
g_2^2\tr{\hC_\mu\hC^\mu} -
\left(I^Z_{111}(1)-I^W_{111}(1)\right)
\frac{1}{2}\frac{1}{\MW^2}\lag_0\right].
\label{eq:P3terms}
\eeqar

Finally, we reintroduce the background Goldstone fields
$\hat{\varphi}_i$ by inverting the Stueckelberg transformation
\refeq{eq:sttrafo}, i.e. we transform the background fields
$\hW_\mu$ and $\hB_\mu$ as
\beq
\hW^\mu\to\hU^\dagger\hW^\mu\hU+\frac{i}{g_2}\hU^\dagger\partial^\mu\hU,
\qquad \hB^\mu \to \hB^\mu.
\label{eq:stinv}
\eeq
The transformations of the fields, field-strength tensors and derivatives
in the standard traces \refeq{eq:sttr} under the Stueckelberg
transformation \refeq{eq:stinv} are given by
\beqar
&&\hC^\mu\to \frac{i}{g_2}\hU^\dagger\hV^\mu\hU, \qquad
\hD^\mu_W\hC_\mu \to
\frac{i}{g_2}\hU^\dagger\left(\hD^\mu_W\hV_\mu\right)\hU,\nn\\
&&\rlap{$\hW^{\mu\nu}\to\hU^\dagger\hW^{\mu\nu}\hU, \qquad$}
\phantom{\hC^\mu\to \frac{i}{g_2}\hU^\dagger\hV^\mu\hU, \qquad}
\hB^{\mu\nu}\to \hB^{\mu\nu}.
\eeqar
Consequently, the traces \refeq{eq:sttr} take
their gauge-invariant form (lhs of the arrow in \refeq{eq:sttr}).
Collecting all terms, we find
\beqar
\leff&=\fac\Bigg\{&\spc g_2\frac{3\MH^2}{4\MW}I_{010} \hH
+g_2^2\left(
\frac{3\MH^2 }{16\MW^2} I_{010}+\frac{9\MH^4}{16\MW^2}
I_{020}\right)\hH^2\nl
+g_2
\left(\frac{3\MH^2}{8\MW} I_{020} -\frac{3\MH^2}{2\MW}
I_{121}\right)\hH\tr{\hV_\mu \hV^\mu}\nl
+\Bigg(\frac{1}{4}I_{010}
+\MW^2 I_{011}-
I^W_{111}(1)\Bigg)\tr{\hV_\mu \hV^\mu}\nl
+g_2^2\bigg(-\frac{1}{2}I_{112} +2
I_{213}\bigg)\tr{\hW_{\mu\nu}\hW^{\mu\nu}}
+ g_1^2\bigg(-\frac{1}{4}I_{112} +
I_{213}\bigg)\hB_{\mu\nu}\hB^{\mu\nu}\nl
+ \bigg(\frac{1}{2}\frac{g_1^2}{g_2^2}I_{011}
+\frac{1}{2\MW^2}\left[I^W_{111}(1)-I^Z_{111}(1)\right]\bigg)\lag_0\nl
+\frac{g_1}{g_2}\bigg(-I_{112}+4I_{213}\bigg)\lag_1
+\frac{g_1}{g_2}\bigg(-\frac{1}{2}I_{112}+4I_{213}\bigg)
\lag_2
+\bigg(\frac{1}{2}I_{112}-4I_{213}\bigg)\lag_3\nl
+\bigg(-4I_{213}+2I_{222}\bigg)\lag_4
+\bigg(\frac{1}{16}I_{020}-\frac{1}{2}I_{121}
+4I_{213}+I_{222}\bigg)\lag_5\nl
+\bigg(-I_{011}+5I_{112}-12I_{213}\bigg)\lag_{11}\Bigg\}\nl
\back+\ord{-2}.
\label{eq:leff2}
\eeqar
This Lagrangian is manifestly invariant under the gauge
transformations of the background fields \refeq{eq:bgtrafo}, under
which the quantities occurring  in \refeq{eq:leff2} with \refeq{eq:sttr}
transform covariantly according to
\beqar
&&\rlap{$\hW^{\mu\nu}\to S \hW^{\mu\nu}S^\dagger$,}
\phantom{\hV^\mu\to S\hV^\mu S^\dagger,}
\qquad\quad\hB^{\mu\nu}\to \hB^{\mu\nu}\nn\\
&&\hV^\mu\to S\hV^\mu S^\dagger,
\qquad\quad\hD^\mu_W\hV_\mu \to S
\left(\hD^\mu_W\hV_\mu\right)S^\dagger,\qquad
\hT\to S\hT S^\dagger.
\eeqar
The gauge for the background fields can now be fixed arbitrarily.

\section{Renormalization}
\label{sec:ren}

In the previous sections we have
dealt with bare parameters and bare fields only.
In the following, these bare quantities are marked by a subscript
``0''.  We apply the renormalization transformation to the parameters
\beqar
e     &\to &          e_0 = (1+\de Z_e)e, \nn\\
M_a^2 &\to & {M_a^2}_{,0} = M_a^2 + \delta M_a^2,
\qquad a=W, Z, H,\nn\\
t     &\to &          t_0 = t + \delta t.
\label{eq:parren}
\eeqar
The tadpole term $t=v(\mu^2-\lambda v^2/4)$ is
defined in the Lagrangian \refeq{eq:smlag} via the term
$tH(x)$.

We apply on-shell renormalization
\cite{bfm5,de93}, where $\MW$, $\MZ$ and $\MH$ represent the physical
masses (propagator poles). The electric unit charge is defined in the
Thomson limit as usual, and the renormalized tadpole vanishes%
\footnote{This means that the relation \refeq{eq:vev} holds for {\em
renormalized\/} quantities, whereas for unrenormalized parameters
$t_0$-terms occur. In order to avoid confusion we omitted $t$ in the
previous sections, but reintroduce it here.}
($t=0$).
The remaining renormalized parameters are
fixed by the relations
\beq
\cw=\frac{\MW}{\MZ}, \qquad \sw=\sqrt{1-\cw^2}, \qquad
g_1=\frac{e}{\cw},\qquad g_2=\frac{e}{\sw},\qquad v=\frac{2\MW}{g_2},
\qquad \mu^2=\frac{\MH^2}{2}.
\eeq
The on-shell conditions imply for the counterterms in
\refeq{eq:parren}
\beqar
\de M_a^2 &=& \Re\left\{\Si^{\hat{a}\hat{a}}_{\mathrm T}(M_a^2)\right\},
\qquad a= W, Z, \nn\\
\de \MH^2 &=& \Re\left\{\Si^{\hH\hH}(\MH^2)\right\}, \nn\\
\de Z_e  &=&
\left.\frac{1}{2}\frac{\partial\Si_{\mathrm T}^{\hA\hA}(q^2)}{\partial
q^2}\right|_{q^2=0},\nn\\
\de  t&=& - T^{\hH},
\label{eq:ctos}
\eeqar
where $\Si_{\mathrm T}^{\hA\hA}$, $\Si_{\mathrm T}^{\hW\hW}$,
$\Si_{\mathrm
T}^{\hZ\hZ}$and $\Si^{\hH\hH}$ represent
the transversal parts of the unrenormalized
vector-boson self-energies and the
unrenormalized $\hH$--self-energy, respectively\footnote{Note that
$\de Z_e$ gets no contribution from the $\hA\hZ$-mixing self-energy
owing to $\Sigma_T^{\hA\hZ}(0)=0$, which follows from the Ward
identity $\Sigma_L^{\hA\hZ}(q^2)=0$ \cite{bfm4,bfm5}.}.
Concerning vertex functions and self-energies our notation follows the
one of
\citeres{bfm4,bfm5} throughout.
Since $\de Z_e$, $\de \MW^2$, $\de \MZ^2$ and $\de t$ are calculated
from vertex functions at low-energy scales, i.e. $|q^2|\ll\MH^2$,
they can be read directly
from the effective Lagrangian \refeq{eq:leff2}, which is constructed
at $|q^2|\ll\MH^2$. However, $\de\MH^2$ is fixed at $q^2=\MH^2$ and
thus cannot be read from \refeq{eq:leff2} but has to be calculated
diagrammatically. As it turns out below,
$\de\MH^2$ is only needed at $\OH{4}$ so that we merely
have to consider those diagrams contributing to the $\hH$--self-energy,
which have internal Higgs or Goldstone lines but no vector lines,
as shown in \reffi{dmhdiag}.
\begin{figure}
\begin{center}
\begin{picture}(16,2.5)
\put(-3.8,-14.8){\includegraphics{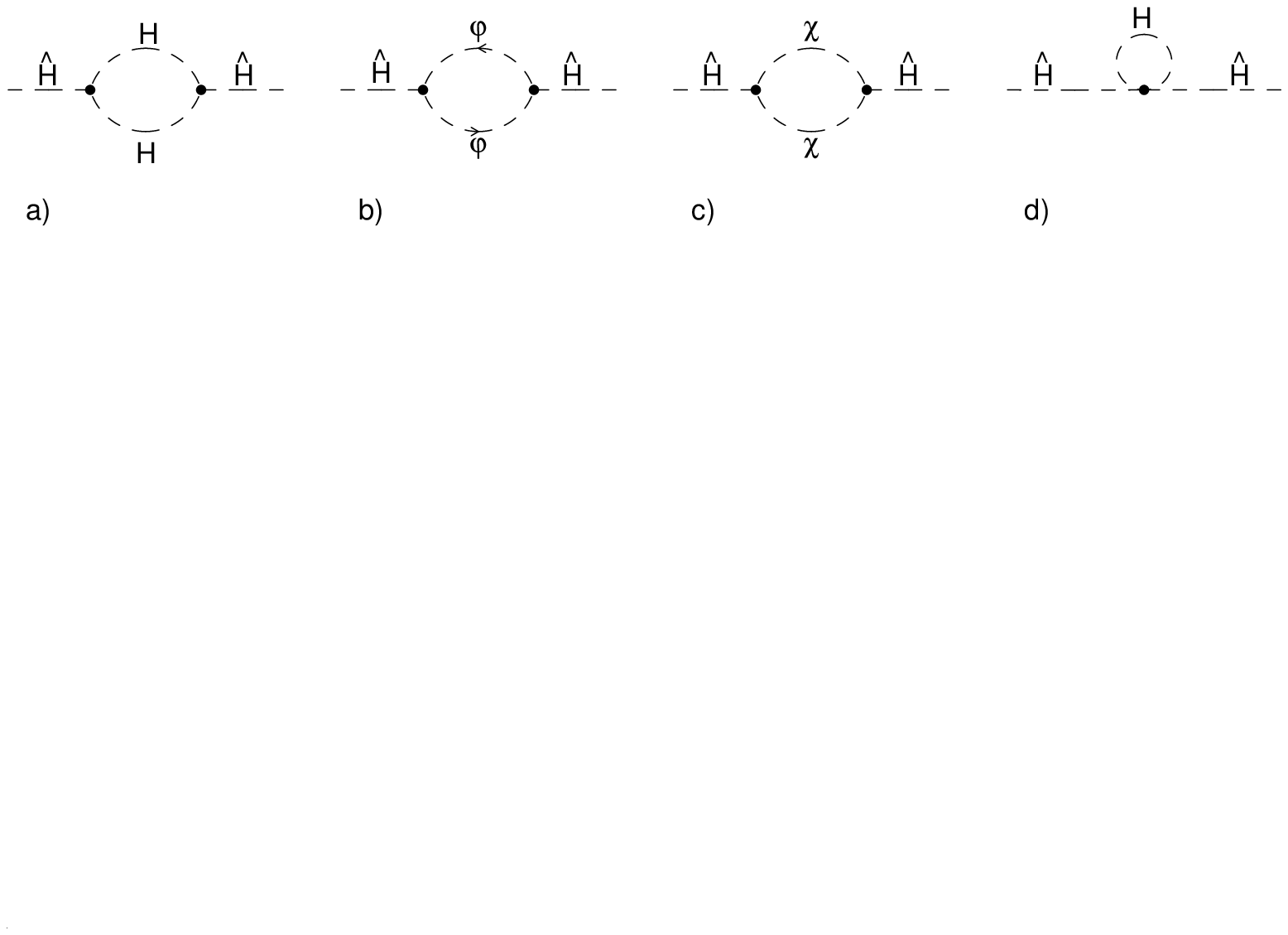}}
\end{picture}
\end{center}
\caption{All diagrams of $\O(\MH^4)$ contributing to $\de\MH^2$.}
\label{dmhdiag}
\efi
We find
\beqar
\de\MH^2 &=& \frac{1}{16\pi^2}g_2^2\,
\frac{3\MH^2}{8\MW^2} \left[
\MH^2\Re\left\{B_0(\MH^2,0,0)\right\}
+3\MH^2 B_0(\MH^2,\MH,\MH)
+I_{010}
\right]+\O(\MH^2),\nn\\
\de\MW^2 &=& \fac g_2^2\left(\frac{1}{4}I_{010}-I^W_{111}(1)\right)
+\O(\MH^0), \nn\\
\de\MZ^2 &=& \frac{\MZ^2}{\MW^2}\de\MW^2+\OH{0},\nn\\
\de t &=& -\fac g_2 \frac{3\MH^2}{4\MW}I_{010} + \O(\MH^0), \nn\\
\de Z_e &=& \O(\MH^0),
\label{eq:dmh}
\eeqar
where $B_0$ denotes the general scalar two-point function
\beq
B_0(k^2,M_0,M_1)=
\frac{(2\pi\mu)^{4-D}}{i\pi^2}\int d^D p\;
\frac{1} {[p^2-M_0^2+i\varepsilon] [(p+k)^2-M_1^2+i\varepsilon]}.
\label{eq:b0}
\eeq
The $B_0$-terms occurring in \refeq{eq:dmh}
are
explicitly
given in \refapp{app:ints}.

In addition we introduce the field renormalization
\beq
\hF \to \hF_{0} = Z_{\hF}^{1/2} \hF
= (1+\frac{1}{2}\de Z_{\hF})\hF,
\qquad F= W, B, H, \vp.
\label{eq:fieren}
\eeq
The renormalized Lagrangian remains gauge-invariant \cite{bfm5},
if one chooses
\beq
\de Z_{\hW}=-2\frac{\de g_2}{g_2},\qquad
\de Z_{\hB}=-2\frac{\de g_1}{g_1},\qquad
\de Z_{\hvp} = 2\frac{\de v}{v},
\label{eq:zwzp}
\eeq
while $\de Z_{\hH}$ can be chosen arbitrarily.
Since $\de Z_{\hH}$ drops out anyhow when $\hH$ is removed from the
theory, we can simply choose
\beq
\de Z_{\hH}=0.
\label{eq:zh}
\eeq
With the choice \refeq{eq:zwzp} the propagators of the massive gauge bosons
acquire residues different from one. However, for the construction of the
effective Lagrangian we only need for the gauge-boson
field-renormalization constants that $\de Z_{\hW}$ and $\de Z_{\hB}$
only
get contributions of $\OH{0}$. This means that we could equivalently
well normalize the residues of all gauge-boson propagators to one
without affecting the final result of the effective Lagrangian.
On the other hand, the condition \refeq{eq:zwzp} for $\de Z_{\hvp}$ is
indeed necessary, because it guarantees that the renormalization of the
matrix $\hU$ \refeq{eq:phinl} does not yield contributions of $\OH{2}$.

As discussed in \citere{sdcgk},
we do not have to carry out the complete renormalization
for the calculation of the effective Lagrangian.
It is sufficient to determine
the $\hH$-dependent part of the counterterm Lagrangian
\beq
\lcth =
\de t\hH - \frac{1}{2}\de\MH^2\hH^2
- \frac{1}{2}\frac{\de\MW ^2}{g_2\MW}\hH\tr{\hV^\mu\hV_\mu} +
\O(\zeta^{-2}).
\label{eq:lcth}
\eeq
This part yields contributions when
eliminating the background field $\hH$ in the next section, i.e.\ in
a diagrammatical procedure these terms contribute to reducible
diagrams with internal Higgs tree lines. Therefore, we do not have to
calculate the counterterms completely, but only those contributions
which yield $\OH{0}$ effects to the final Lagrangian. In particular,
$\de \MH^2$ only has to be determined at $\OH{4}$, because
$\hH$ turns out to be
$\OH{-2}$ when it will be eliminated in the next section. For the same
reason it is sufficient to consider $\de\MW^2$
only at $\OH{2}$.

As in \citere{sdcgk},
we call the sum of $\leff$ \refeq{eq:leff2} and
$\lcth$ \refeq{eq:lcth} the {\em renormalized effective Lagrangian\/}
$\lreff$.
Inserting \refeq{eq:dmh},
we find for the $\hH$-dependent part of $\lreff$
\beqar
\lreff\vert_{\hH}&=
\fac\Bigg\{&\spc \frac{3g_2^2\MH^4}{16\MW^2} \bigg(
3I_{020}-3B_0(\MH^2,\MH,\MH)-\Re\left\{B_0(\MH^2,0,0)\right\}
\bigg)\hH^2
\nl
+ \frac{g_2}{8\MW}
\bigg(-I_{010}+4I^W_{111}(1)+3\MH^2I_{020}
-12\MH^2I_{121}\bigg)\hH\tr{\hV_\mu \hV^\mu}\Bigg\}
\nl
\back+\ord{-2},
\label{eq:leff3}
\eeqar
while the $\hH$-independent part is obviously the same as in
\refeq{eq:leff2}.

\section{Elimination of the background Higgs field}
\label{sec:bghelim}

Having integrated out the quantum Higgs field $H$, which corresponds
to Higgs lines in loops, the effective Lagrangian still
contains the background Higgs field $\hH$, which corresponds to
Higgs tree lines in Feynman diagrams. The field $\hH$
can now be eliminated in complete analogy to the
procedure of \citere{sdcgk} so that we discuss this point only briefly here.
Since the $\hH$-field corresponds to tree
lines, the $\hH$-propagators can be expanded in powers of $1/\MH^2$
for $\MH\to\infty$.
Diagrammatically this means that the $\hH$-propagator shrinks to a point
rendering such \mbox{(sub-)}graphs irreducible which contain $\hH$-lines
only.
The tree-level Lagrangian of the SM
implies that this expansion corresponds to
the replacement
\beq
\hH\to -\frac{\MW}{g_2\MH^2}\tr{\hV_\mu\hV^\mu} +\OH{-4}.
\label{eq:heomex}
\eeq
The
substitution
\refeq{eq:heomex} can be alternatively motivated by the
fact that it corresponds to the use of the
equation of motion (EOM)
for the
background Higgs field,
which is fulfilled in lowest order by the
tree-like part of Feynman diagrams. After applying \refeq{eq:heomex},
the effective Lagrangian $\lreff$ becomes:
\beqar
\lreff&=
\fac\Bigg\{&\spc
\bigg(\frac{1}{4}I_{010}
+\MW^2 I_{011}-
I^W_{111}(1)\bigg)\tr{\hV_\mu \hV^\mu}\nl
+ g_2^2\left(-\frac{1}{2}I_{112} +2
I_{213}\right)\tr{\hW_{\mu\nu}\hW^{\mu\nu}}
\nl
+ g_1^2\bigg(-\frac{1}{4}I_{112} +
I_{213}\bigg)\hB_{\mu\nu}\hB^{\mu\nu}\nl
+ \bigg(\frac{1}{2}\frac{g_1^2}{g_2^2}I_{011}
+\frac{1}{2\MW^2}\left[I^W_{111}(1)-I^Z_{111}(1)\right]\bigg)\lag_0\nl
+\frac{g_1}{g_2}\bigg(-I_{112}+4I_{213}\bigg)\lag_1\nl
+\frac{g_1}{g_2}\bigg(-\frac{1}{2}I_{112}+4I_{213}\bigg)
\lag_2\nl
+\bigg(\frac{1}{2}I_{112}-4I_{213}\bigg)\lag_3\nl
+\bigg(-4I_{213}+2I_{222}\bigg)\lag_4\nl
+\bigg(\frac{1}{8\MH^2}I_{010}-\frac{1}{2\MH^2}I_{111}(1)
+\frac{1}{4}I_{020}+I_{121}
+4I_{213}+
I_{222}
\nl\qquad\qquad\qquad
-\frac{9}{16}B_0(\MH^2,\MH,\MH)
-\frac{3}{16}\Re\left\{B_0(\MH^2,0,0)\right\}
\bigg)\lag_5\nl
+\bigg(-I_{011}+5I_{112}-12I_{213}\bigg)\lag_{11}\Bigg\}\nl
\back+\OH{-2}.
\label{eq:lefffinal}
\eeqar
Finally, we insert the explicit forms \refeq{integrals} and
\refeq{Bs} of the integrals in this expression and find:
\beqar
\lreff&=\fac\Bigg\{&
\spc \bigg[-\frac{1}{8}\MH^2+\frac{3}{4}\MW^2\left(\tde
+\frac{5}{6}\right)
\bigg]\tr{\hV_\mu \hV^\mu}\nl
-\frac{1}{24}
\bigg(\tde+\frac{5}{6}\bigg)g_2^2\tr{\hW_{\mu\nu}\hW^{\mu\nu}}
\nl
- \frac{1}{48}\bigg(\tde+\frac{5}{6}\bigg)g_1^2
\hB_{\mu\nu}\hB^{\mu\nu}\nl
+ \frac{3}{8}\bigg(\tde+\frac{5}{6}\bigg)\frac{g_1^2}{g_2^2}\lag_0\nl
-\frac{1}{12}\bigg(\tde+\frac{5}{6}\bigg)\frac{g_1}{g_2}\lag_1\nl
+\frac{1}{24}
\bigg(\tde+\frac{17}{6}\bigg)
\frac{g_1}{g_2}\lag_2\nl
-\frac{1}{24}\bigg(\tde+\frac{17}{6}\bigg)\lag_3\nl
-\frac{1}{12}\bigg(\tde+\frac{17}{6}\bigg)\lag_4\nl
-\frac{1}{24}\bigg(\tde+\frac{79}{3}-\frac{9\sqrt{3}\pi}{2}\bigg)
\lag_5\nl
-\frac{1}{4}\bigg(\tde+\frac{1}{6}\bigg)\lag_{11}\Bigg\}\nl
\back+\OH{-2},
\label{eq:lefffinalex}
\eeqar
with $\tde$ being given in \refeq{tde}.

The tree-level Lagrangian of the
SM for $\MH\to\infty$ is the Lagrangian of the corresponding
$\mathrm{SU(2)_{\mathrm W}\times U(1)_{\mathrm Y}}$
gauged non-linear $\sigma$-model (GNLSM) \cite{apbe,long}, which is
obtained from the SM Lagrangian simply be dropping the Higgs field
in the non-linear realization of the scalar fields \refeq{eq:smlag2}
\beq
\lag^{\mathrm{tree}}|_{\MH\to\infty}=\lag^{\mathrm{tree}}|_{\hH=0}+\OH{-2}
=\lag^{\mathrm{tree}}_{\mathrm{GNLSM}}+\OH{-2},
\label{eq:ltree}
\eeq
with
\beq
\lag^{\mathrm{tree}}_{\mathrm{GNLSM}}=
-\frac{1}{2} \tr{\hW_{\mu\nu}\hW^{\mu\nu}}-\frac{1}{4}\hB_{\mu\nu}
\hB^{\mu\nu}
-\frac{\MW^2}{g_2^2}\tr{\hV_\mu \hV^\mu }.
\label{eq:lgnlsmtree}
\eeq
The complete one-loop Lagrangian
$\left.\lag^{\mathrm{1-loop,ren}}\right|_{\MH\to\infty}$
of the SM for $\MH\to\infty$ consists of three different parts:
The effective Lagrangian $\lreff$, the part
$\left.\lag^{\mathrm{1-loop}}\right|_{H=0}$ of the one-loop
Lagrangian which does not contain the quantum Higgs field $H$,
and the part $\left.\lag^{\mathrm{ct}}\right|_{\hH=0}$
of the countertem Lagrangian which does not contain the background
field $\hH$. As in \citere{sdcgk}, one can easily show
that eliminating the background Higgs field $\hH$
in $\left.\lag^{\mathrm{1-loop}}\right|_{H=0}$
by applying \refeq{eq:heomex} simply results in
dropping all terms which contain $\hH$. Thus,
we find that
the one-loop Lagrangian of the SM for $\MH\to\infty$ is the
sum of the one-loop Lagrangian of the GNLSM, the corresponding
counterterm Lagrangian, and the effective Lagrangian
\beqar
\lag^{\mathrm{1-loop,ren}}|_{\MH\to\infty} &=&
\lag^{\mathrm{1-loop}}|_{H=\hH=0}+
\lag^{\mathrm{ct}}|_{\hH=0}+
\lreff+\OH{-2} \nn\\[.3em]
&=& \lag^{\mathrm{1-loop}}_{\mathrm{GNLSM}}+
\lag^{\mathrm{ct}}_{\mathrm{GNLSM}}+\lreff+\OH{-2}.
\label{eq:lfinal}
\eeqar
The counterterm Lagrangian $\lag^{\mathrm{ct}}_{\mathrm{GNLSM}}$
follows from the
tree-level Lagrangian of the GNLSM \refeq{eq:lgnlsmtree} by applying
the renormalization transformations \refeq{eq:parren} and
\refeq{eq:fieren}. The renormalization constants occurring in
$\lag^{\mathrm{ct}}_{\mathrm{GNLSM}}$
are calculated from self-energies, as e.g. given in
\refeq{eq:ctos} for the mass and charge renormalization constants.
Of course, the contribution of the effective Lagrangian $\lreff$ to
the relevant self-energies have to be included in this procedure.

The first three terms in \refeq{eq:lefffinalex} have the same
structure as terms in the tree-level Lagrangian of the GNLSM
\refeq{eq:lgnlsmtree}. They can be absorbed into the corresponding
counterterms and have no effect on S-matrix elements.
Furthermore, the $\lag_{11}$-term in \refeq{eq:lefffinalex}
does not affect S-matrix elements%
\footnote{$\lag_{11}$ yields contributions to S-matrix
elements if massive fermions are included. This is discussed in
the next section.}, because $\lag_{11}$ \refeq{eq:sttr}
can be eliminated by applying the EOMs \cite{eom} for the
SU(2)$_{\PW}$ background vector fields within the GNLSM \cite{sdcgk},
\beq
\DW^\mu \hW_{\mu\nu} = -\frac{i}{g_2}\MW^2\hV_\nu.
\label{eq:DVEOM}
\eeq
Using $\DW^\mu \DW^\nu \hW_{\mu\nu}=0$, this leads to
\beq
\DW^\mu \hV_\mu=0,
\label{eq:DVEOM2}
\eeq
which is
valid at tree-level. Since $\lreff$ only contains
background fields (corresponding to tree lines), this is
sufficient to render the contribution of $\lag_{11}$ to the
S-matrix
zero. Thus, the complete one-loop effects of a heavy Higgs
boson on S-matrix elements, i.e.\ the complete difference between
the SM for $\MH\to\infty$ and the GNLSM
contributing to the S-matrix at one loop, are summarized
in the effective Lagrangian
\beqar
\lreff(\mbox{S-matrix})&=\fac\Bigg\{&
\spc \frac{3}{8}\bigg(\tde+\frac{5}{6}\bigg)\frac{g_1^2}{g_2^2}
{\MW^2}\left(\tr{\hT \hV_\mu}\right)^2\nl
-\frac{1}{24}\bigg(\tde+\frac{5}{6}\bigg)g_1g_2
\hB_{\mu\nu}\tr{\hT\hW^{\mu\nu}}\nl
+\frac{1}{48}
\bigg(\tde+\frac{17}{6}\bigg)
i{g_1}\hB_{\mu\nu}\tr{\hT[\hV^\mu,\hV^\nu]}\nl
-\frac{1}{24}\bigg(\tde+\frac{17}{6}\bigg)ig_2
\tr{\hW_{\mu\nu}[\hV^\mu,\hV^\nu]}\nl
-\frac{1}{12}\bigg(\tde+\frac{17}{6}\bigg)
\left(\tr{\hV_\mu\hV_\nu}\right)^2\nl
-\frac{1}{24}\bigg(\tde+\frac{79}{3}-\frac{9\sqrt{3}\pi}{2}\bigg)
\left(\tr{\hV_\mu\hV^\mu}\right)^2\Bigg\}\nl
\back+\OH{-2},
\label{eq:finalresult}
\eeqar
where the explicit form of the traces \refeq{eq:sttr}
is inserted.

Finally, we note that the result of our functional calculation
\refeq{eq:finalresult} coincides with the result of the diagrammatical
calculation in \citere{hemo}%
\footnote{We find a coefficient
for the $\lag_{11}$-term in \refeq{eq:lefffinalex}
which is different from the one in
\citere{hemo}. This is due to the fact that we use the non-linear
parametrization of the Higgs sector \refeq{eq:phinl} while in \citere{hemo}
the linear one \refeq{eq:philin} is used.
Such a reparametrization of the scalar
fields may change Green functions but not S-matrix elements
\cite{stue2,stue3}. As
pointed out, the $\lag_{11}$-term has no impact on S-matrix
elements (as far as one considers the pure bosonic sector).}.
(Note that our coupling constants $g_1$ and $g_2$
correspond to the constants $g^\prime$ and $g$ in \citere{hemo}
by the substitutions $g_1 \to g^\prime$, $g_2 \to - g$.)

\section{Fermionic contributions to the effective Lagrangian}
\label{sec:fermions}

\subsection{The fermionic part of the standard model Lagrangian}
\label{ssec:bfm2}
In the previous sections we have only considered the bosonic sector of
the electroweak SM. Now, we also include fermions in
our calculation and determine the fermionic terms of the low-energy
effective Lagrangian generated by integrating out the Higgs field.

The fermionic part of the SM Lagrangian is
\beq
\lag_{F}=i\left(\overline{\Psi}_f\Slash{D}_{f,\sigma}\om_\sigma \Psi_f
\right)
-\frac{\sqrt{2}}{v}\left(\overline{\Psi}_f\Mf\Phi^\dagger\om_-\Psi_f
+\overline{\Psi}_f
\Phi\Mf\om_{+}\Psi_f\right),
\label{eq:lagfermi}
\eeq
where the index $f$ labels the different fermion doublets $\Psi_f$
with the mass matrix%
\footnote{We neglect quark mixing throughout, i.e.\ the CKM matrix is
set to the unit matrix. The generalization to finite quark mixing is
straightforward.}
$\Mf$, and $\om_{\pm}$ denote the chirality projectors,
\beq
\Psi_f=\left(\psi_{f_1}\atop\psi_{f_2}\right),\qquad \Mf=\left(
\begin{array}{cc}m_{\Pf_1} & 0\\ 0 & m_{\Pf_2}\end{array}\right), \qquad
\om_\pm= \frac{1}{2} \left(1\pm \gamma_5 \right).
\label{eq:doublet}
\eeq
In \refeq{eq:lagfermi} and the following summation over all doublets
$\Psi_f$ is assumed.
The covariant derivatives are
\beq
D^{\mu}_{f,\sigma} = \partial^\mu -ig_2 W^\mu \delta_{\sigma -}
+\frac{i}{2}g_1Y_{\Pf,\sigma}B^\mu
\label{eq:dcovfermi}
\eeq
with
\beq
Y_{\Pf,\sigma} = 2Q_\Pf - \tau_3 \delta_{\sigma -}
\label{eq:hcf}
\eeq
where $Q_\Pf$ is the electric
charge
matrix
of $\Psi_f$, and $Y_{\Pf,\sigma}$
the weak hypercharge
matrix
of $\om_\si\Psi_f$.
The scalar field $\Phi$ is again non-linearly realized
according to \refeq{eq:phinl}.

The BFM is applied by
splitting the fermion fields linearly according to
\beq
\Psif \to \hp +\Psif, \qquad \overline{\Psi}_f \to \hpb + {\Psibf},
\label{eq:splitfermi}
\eeq
and the boson fields according to \refeq{eq:split}.
Finally, the Stueckelberg transformation of the fermion fields
\cite{stue4,stue3}
\beqar
&&\omm \hp \to \hU \omm \hp, \qquad \omm \Psif \to \hU \omm \Psif, \qquad
\hpb \omp \to  \hpb \omp \hU^\dagger, \qquad
{\Psibf} \omp \to  {\Psibf} \omp \hU^\dagger,\nn\\
&&
\rlap{$\omp \hp \to \omp \hp$,} \,\phantom{\omm \hp \to \hU \omm \hp,}
\qquad \rlap{$\omp \Psif \to \omp \Psif$,}\,
\phantom{\omm \Psif \to \hU \omm \Psif,}\qquad
\rlap{$\hpb \omm \to  \hpb \omm$,}\, \phantom{\hpb \omp \to  \hpb \omp
\hU^\dagger,} \qquad
\rlap{${\Psibf}\omm \to  {\Psibf} \omm$}\nn\\&&
\label{eq:sttrafofermi}
\eeqar
together with
the one
of the bosons \refeq{eq:sttrafo} removes the
background Goldstone fields from the Lagrangian.

\subsection{Diagonalization}
\label{ssec:diag}

The one-loop part of Lagrangian \refeq{eq:lagfermi}
can be written in the symbolic form
\beqar
\lag_{F}^{\mathrm{1-loop}}&=&{}\phantom{+}\,\,\,{\Psibf}\Delta_f\Psif
-\tr{\vp\delta\Delta_{\vp}\vp}+H\tr{\delta X_{\ss H\vp}\vp}
+H{\Psibf}X_{\ss fH}+ H \overline{X}_{\ss fH} \Psif\nl
+{\Psibf}
\Mf\vp\omm X^L_{\ss f\vp}+ \overline{X}_{\ss f\vp}^L\omp\vp \Mf
\Psif + {\Psibf} \vp \Mf \omp X^R_{\ss f\vp} +\overline{X}^R_{\ss f \vp}
\omm\Mf \vp  \Psif\nl
+{\Psibf}\Slash{W}\,
X_{\ss fW}+ \overline{X}_{\ss fW} \Slash{W}\,\Psif+
{\Psibf}\Slash{B}\,X_{\ss fB}+ \overline{X}_{\ss fB}\Slash{B}\,\Psif,
\label{eq:lfsymb}
\eeqar
with the operators
\beqar
\Delta_{f}&=& i\Slash{\hat{D}}_{f,\sigma}\om_\sigma-\Mf
\left(1+\frac{\hH}{v}\right),\nn\\
\delta\Delta_\vp &=& -\frac{g_2^2}{4\MW^2}\hpb\Mf\hp
\left(1+\frac{\hH}{v}\right),\nn\\
\delta X_{\ss H\vp}^{ab} &=& -i \frac{g_2^2}{2\MW^2}
\left[\hpb^b\omp\left(\Mf\hp\right)^a-
\left(\hpb\Mf\right)^b\omm\hp^a\right],\nn\\
X_{\ss fH} &=& -\frac{g_2}{2\MW} \Mf \hp,\nn\\
X_{\ss f\vp}^L&=& i\frac{g_2}{\MW}\hp\left(1+\frac{\hH}{v}\right), \qquad
X_{\ss f\vp}^R\,=\, -i\frac{g_2}{\MW}\hp\left(1+\frac{\hH}{v}\right),\nn\\
X_{\ss fW}&=&g_2\omm\hp,\qquad
X_{\ss fB}\,=\,-\frac{g_1}{2}Y_{\Pf,\sigma}\om_\sigma\hp.
\label{eq:deltasf}
\eeqar
The indices $a$ and $b$ in the third line denote the
SU(2)$_\PW$ indices
of the 2$\times$2-matrix $\delta X_{\ss H\vp}$.

As in \refse{sec:diag}, the mixings between the quantum
Higgs field $H$ and the other quantum fields can be removed by
appropriate shifts of the quantum fields. It turns out to be useful
first to remove the $H\Psif$-mixing in \refeq{eq:lfsymb} before
diagonalizing the bosonic sector of the SM Lagrangian \refeq{eq:lag2}.
This can be achieved by the shifts
\beq
\Psif \to \Psif - {\Delta}_f^{-1}X_{\ss fH}H,\qquad
{\Psibf} \to {\Psibf} - H \overline{X}_{\ss fH}
\overline{{\Delta}_f^{-1}}
\label{eq:psishift}
\eeq
with
\beq
\overline{\Delta_f^{-1}}=
\gamma_0\left({\Delta}_f^{-1}\right)^\dagger\gamma_0,
\eeq
which modify the term bilinear in $H$ and the
$H\vp$-terms in \refeq{eq:lag2} and \refeq{eq:lfsymb} according to
\beq
\Delta_H \to \Delta_H + \delta\Delta_H, \qquad
X_{\ss H\vp} + \delta X_{\ss H\vp} \to
X_{\ss H\vp} + \delta X_{\ss H\vp} + \delta X^\prime_{\ss H\vp}
\label{eq:fshifts}
\eeq
with
\beqar
\delta\Delta_H&=&2
\overline{X}_{\ss fH}{\Delta}_f^{-1} X_{\ss fH}\nn\\
H\tr{\delta X^\prime_{\ss H\vp}\vp}&=&{}
-H\bX_{\ss fH}\left(\Delta_f^{-1} \Mf \vp \omm X_{\ss f\vp}^L\right)
-H\bX_{\ss fH}\left(\Delta_f^{-1} \vp \Mf \omp X_{\ss f\vp}^R\right)\nl
-\bX_{\ss f\vp}^L\omp  \vp \Mf \left(\Delta^{-1}_f X_{\ss fH}H\right)
-\bX_{\ss f\vp}^R\omm \Mf  \vp \left(\Delta^{-1}_f X_{\ss fH}H\right).
\label{eq:deltaxf}
\eeqar
In \refeq{eq:deltaxf}, we define $\delta X^\prime_{\ss H\vp}$ implicitly
via $H\tr{\delta X^\prime_{\ss H\vp}\vp}$ since its explicit expression
outside the trace is not needed in the following.
In addition to \refeq{eq:fshifts},
there is a modification of the $HW$- and $HB$-terms, which
however can be neglected at $\OH{0}$. We also had to remove the $f\vp$-,
$fW$- and $fB$-terms by appropriate shifts before doing the shifts
\refeq{eq:shifts} in the bosonic sector (such that those do not effect
the fermionic sector), and
finally reverse these shifts in order to restore these terms.
However, it turns out by simple power counting that the contributions
of these shifts to the $\Delta$s and $X$s in the bosonic sector
\refeq{eq:DelX}  only yield $\OH{-2}$-effects.

This means that all fermionic $\OH{0}$-contributions to
$\leff$ can be found by adding $\delta\Delta_H$, $\delta\Delta_\vp$,
$\delta X_{\ss H\vp}$ and $\delta X_{\ss H\vp}^\prime$ given by
\refeq{eq:deltasf} and \refeq{eq:deltaxf} to the bosonic parameters
\refeq{eq:DelX}, and
proceeding as in the calculation of the
bosonic part of $\leff$. Thus, $\tthn$ \refeq{eq:ttDH} modifies to
\beqar
\tthn&\to&
\tthn+\delta\tthn\nn\\
&=&\tthn+
\delta\Delta_H-\frac{1}{2}\tr{X_{\ss H\vp}
\delta\left(\hat{\Delta}_\vp^{-1}\right)X_{\ss H\vp}^\dagger}
-\frac{1}{2}\left(\tr{\delta X_{\ss H\vp} \hat{\Delta}_\vp^{-1}
X_{\ss H\vp}^\dagger}+\mbox{h.c.}\right)\nl
-\frac{1}{2}\tr{\delta X_{\ss H\vp} \hat{\Delta}_\vp^{-1}
\delta X_{\ss H\vp}^\dagger}
-\frac{1}{2}\left(\tr{\delta X^\prime_{\ss H\vp} \hat{\Delta}_\vp^{-1}
X_{\ss H\vp}^\dagger}+\mbox{h.c.}\right)+\ord{-3}
\label{eq:ttDHf}
\eeqar
with
\beq
\delta\left(\hat{\Delta}_\vp^{-1}\right)=\widehat{(\Delta_\vp+
\delta\Delta_\vp)}^{-1}-\hat{\Delta}_\vp^{-1}.
\eeq
In \refeq{eq:ttDHf} terms yielding only $\OH{-2}$-contributions are again
neglected.

\subsection{{\protect\boldmath$1/\MH$}-Expansion}
\label{ssec:exp}
The fermionic part of $\leff$
can be derived by expanding the contribution of
$\delta\tthn$ in \refeq{eq:ttDHf} to \refeq{eq:leff}
in analogy to the procedure described in \refse{sec:helim}. This yields
\beqar
\delta\leff&=\fac\Bigg\{&\spc\frac{g_2^2}{4\MW^2}I_{011}
\hpb\Mf^3\hp
+\frac{i g_2^2}{4\MW^2}\left(I_{011}-2I_{112}\right)
\hpb\Mf\Slash{\hat{D}}_{f,\sigma}\Mf \om_{-\sigma}\hp\nl
-\frac{g_2^3}{2\MW^2}I_{112}\hpb\left[2\Mf\Slash{\hat{C}}\Mf\omp-\left(
\Mf^2\Slash{\hat{C}}+\Slash{\hat{C}}\Mf^2\right)\omm\right]\hp\nl
-\frac{g_2^4}{4\MW^2}I_{112}\hpb\Mf\hp\tr{\hC_\mu\hC^\mu}\nl
-\frac{i g_2^3}{2\MW^2}\left(I_{011}-2I_{112}\right)
\hpb\left[ \left(\hD_W^\mu \hC_\mu\right) \Mf\omp- \Mf\left(\hD_W^\mu
\hC_\mu\right) \omm\right]\hp\nl
-\frac{g_2^4}{32\MW^4}I_{011}\left[\hpb\left(
\tau_i \Mf \omp- \Mf \tau_i \omm\right) \hp\right]
\nl\qquad\qquad\qquad\qquad\times
\left[\hpbp\left(
\tau_i \Mfp \omp- \Mfp \tau_i \omm\right) \hpp\right]\Bigg\}\nl
\back+\ord{-2}.
\label{eq:dleff1}
\eeqar
Strictly speaking, in \refeq{eq:dleff1} vacuum integrals of the form
\beqar
\frac{(2\pi\mu)^{4-D}}{i\pi^2}
\int {d^D p} \frac{\disp
p_{\mu_1}\ldots p_{\mu_{2k}}}{(p^2-\MH^2)^l(p^2-M_1^2)^{m_1}
(p^2-M_2^2)^{m_2}}
\quad \mbox{with} \quad M_{1,2}^2=\xi M_{\PW,\PZ}^2, m_{\Pf_i}^2
&& \hspace{2em}
\label{eq:intnotf}
\eeqar
occur, because in addition to the bosonic propagators there are also
fermionic ones. Since in \refeq{eq:dleff1} only logarithmically
divergent integrals are relevant, which
are independent of $M_1^2$ and $M_2^2$ (and thus
depend only on $m=m_1+m_2$) at $\OH{0}$,
these are still given by the explicit expressions \refeq{integrals}
for the integrals $I_{klm}$ \refeq{eq:intnot}.
In particular, the fact that
the fermion masses within a doublet can be different
does not effect these integrals at $\OH{0}$.

The origin of the various terms in $\leff$ \refeq{eq:dleff1} is the
following: the first two terms are the contribution of
$\delta\Delta_H$ in \refeq{eq:ttDHf},
the third term is the contribution of $\delta X^\prime_{\ss H\vp}
\hat{\Delta}_\vp^{-1} X_{\ss H\vp}^\dagger+\mbox{h.c.}$, the fourth
stems from $X_{\ss H\vp}
\delta\left(\hat{\Delta}_\vp^{-1}\right)X_{\ss H\vp}^\dagger$, the fifth from
$\delta X_{\ss H\vp} \hat{\Delta}_\vp^{-1}
X_{\ss H\vp}^\dagger+\mbox{h.c.}$, and the last from
$\delta X_{\ss H\vp} \hat{\Delta}_\vp^{-1}
\delta X_{\ss H\vp}^\dagger$. Note that the explicit occurrence of
the Pauli matrices $\tau_i$ in the last term in \refeq{eq:dleff1}
is a consequence of the operator $P$ \refeq{eq:P} in
$\hat\Delta_{\vp}^{-1}\dpp$ \refeq{eq:dpinv}.

\subsection{The Stueckelberg formalism}
\label{stue}
We invert the Stueckelberg transformation \refeq{eq:sttrafo},
\refeq{eq:sttrafofermi} in order to rewrite
$\delta\leff$
in a gauge-invariant form. The inverse Stueckelberg
transformation is given by \refeq{eq:stinv} and
\beq
\omm \hp \to \hU^\dagger \omm \hp, \;\;\quad \hpb \omp \to \hpb \omp \hU,
\;\;\quad\omp \hp \to \omp \hp , \;\;\quad \hpb \omm \to \hpb \omm.
\label{eq:stueinvf}
\eeq
This yields
\beqar
\delta\leff&=\fac\Bigg\{&\spc
\frac{g_2^2}{4\MW^2}I_{011}
\hpb\left(\hU\Mf^3\omp+\Mf^3\hUd\omm\right)\hp\nl
+\frac{ig_2^2}{4\MW^2}\left(I_{011}-2I_{112}\right)
\hpb\left(\Mf\hUd\Slash{\hat{D}}_{f,-}\hU\Mf\omp +
\hU\Mf\Slash{\hat{D}}_{f,+}\Mf\hUd\omm
\right) \hp\nl
-\frac{i g_2^2}{2\MW^2}I_{112}
\hpb\left[2\Mf\hUd\Slash{\hat{V}}\hU\Mf\omp-\left(
\hU\Mf^2\hUd\Slash{\hat{V}}+\Slash{\hat{V}}\hU\Mf^2\hUd\right)\omm\right]\hp\nl
+\frac{g_2^2}{4\MW^2}I_{112}\hpb\left(
\hU\Mf\omp+\Mf\hUd\omm \right)\hp\tr{\hV_\mu\hV^\mu}\nl
+\frac{g_2^2}{2\MW^2}\left(I_{011}-2I_{112}\right)
\hpb\left[ \left(\hD_W^\mu \hV_\mu\right)\hU \Mf\omp- \Mf\hUd\left(\hD_W^\mu
\hV_\mu\right) \omm\right]\hp\nl
-\frac{g_2^4}{32\MW^4}I_{011}\left[\hpb\left(\hU
\tau_i \Mf \omp- \Mf \tau_i \hUd \omm\right) \hp\right]
\nl\qquad\qquad\qquad\qquad\times
\left[\hpbp\left(\hU
\tau_i \Mfp \omp- \Mfp \tau_i \hUd \omm\right) \hpp\right]\Bigg\}\nl
\back+\ord{-2}.
\label{eq:dleff2}
\eeqar
This Lagrangian is invariant under the background gauge
transformations \refeq{eq:bgtrafo} and
\beqar
&&\rlap{$\omm\hp\to S S_{Y_{\Pf,-}}\omm\hp,$}
\phantom{\hpb\omp \to \hpb \omp S_{Y_{\Pf,-}}^\dagger S^\dagger}
 \qquad \omp \hp \to  S_{Y_{\Pf,+}}\omp
\hp, \nn\\&&
\hpb\omp \to \hpb \omp S_{Y_{\Pf,-}}^\dagger S^\dagger
,\qquad \hpb\omm\to \hpb\omm S_{Y_{\Pf,+}}^\dagger,
\label{eq:bgtrafof}
\eeqar
where $S$ is given by \refeq{eq:S} and $S_{Y_{\Pf,\sigma}}$ by
\beq
S_{Y_{\Pf,\sigma}}=\exp\left(-\frac{i}{2}g_1Y_{\Pf,\sigma}
\theta_{\mathrm{Y}}\right)
\eeq
with the weak hypercharges $Y_{\Pf,\sigma}$ \refeq{eq:hcf}.

The second term in \refeq{eq:dleff2} can be simplified by applying the product
rule for the covariant derivatives.
This yields a term with derivatives acting only on the fermion fields and
a term which has the same structure as the third term in \refeq{eq:dleff2}.
The dimension-4 part (i.e.\ the
second and third term) of \refeq{eq:dleff2} becomes
\beqar
\left.\delta\leff\right|_{\rm dim=4}&
=\fac\Bigg\{&\spc
\frac{ig_2^2}{8\MW^2}\left(I_{011}-2I_{112}\right)
\left[\hpb\left(\Mf^2\Slash{\hat{D}}_{f,+}\omp +
\hU\Mf^2\hUd\Slash{\hat{D}}_{f,-}\omm
\right) \hp+{\rm h.c.}\right]\nl
+\frac{i g_2^2}{8\MW^2}(I_{011}-6I_{112})
\hpb\Big[2\Mf\hUd\Slash{\hat{V}}\hU\Mf\omp
\nl\qquad\qquad\qquad\qquad-\left(
\hU\Mf^2\hUd\Slash{\hat{V}}+\Slash{\hat{V}}\hU\Mf^2\hUd\right)\omm\Big]\hp
\Bigg\}.
\eeqar

\subsection{Renormalization}
\label{ssec:ren}

In analogy to \refse{sec:ren},
we have to add the fermionic part of
the Higgs dependent counterterms to $\delta\leff$.
The parameter- and field-renormalization transformations
of the fermions are
\beqar
m_{\Pf_i} &\to& m_{\Pf_i,0} = m_{\Pf_i} + \delta m_{\Pf_i}, \nn\\
\om_\si\hat{\psi}_{f_i}&\to&\om_\si\hat{\psi}_{f_i,0}=(Z_{f_i}^\si)^{1/2}
\om_\si\hat{\psi}_{f_i}=(1+\frac{1}{2}\delta Z_{f_i}^\si)
\om_\si\hat{\psi}_{f_i}.
\eeqar
{}From \refeq{eq:dleff2} one immediately reads
\beq
\frac{\delta m_{\Pf_i}}{m_{\Pf_i}}=\OH{0},
\qquad \delta Z_{f_i}^\si= \OH{0}.
\label{eq:fercts}
\eeq
In this context, one should notice that the renormalized effective
action only remains gauge-invariant if the left-handed fermion-doublet
field
$\omm\Psi_f$ is renormalized by {\it one\/} renormalization constant,
i.e.\ $\delta Z_f^L=\delta Z_{f_1}^L=\delta Z_{f_2}^L$ (in $\delta Z_f$
the superscripts R/L are used instead of $\si=+/-$). Similarly to the
case of the gauge-boson fields considered in \refse{sec:ren},
the explicit form of the field-renormalization constants $\delta Z_{f_i}^\si$
is irrelevant for the construction of the effective Lagrangian
as long as \refeq{eq:fercts} holds.
In particular, \refeq{eq:fercts} is fulfilled in the complete on-shell
scheme \cite{de93}, where all fermion propagators acquire residues equal
to one. According to simple power counting,
we only have to consider the contribution of $\delta\MW^2$ to
$\delta\lcth$:
\beq
\delta\lag_{\mathrm{\hH}}^{\mathrm{ct}}=\frac{g_2}{4\MW^3}\delta\MW^2
\hH\hpb\left(\hU\Mf\omp+\Mf\hUd\omm\right)\hp+\ord{-2},
\label{eq:dlct}
\eeq
with $\delta\MW$ given in \refeq{eq:dmh}.
The fermionic part $\delta\lreff$ of the renormalized effective Lagrangian
is the sum of $\delta\leff$ \refeq{eq:dleff2} and $\delta\lcth$
\refeq{eq:dlct}.

\subsection{Elimination of the background Higgs field}
\label{ssec:elim}

As in \refse{sec:bghelim}, we can eliminate the background Higgs
field $\hH$ by a propagator expansion, or equivalently by an
application of the
EOM for $\hH$ in lowest order.
The fermionic part of the SM Lagrangian \refeq{eq:lagfermi} implies that
\refeq{eq:heomex} generalizes to
\beq
\hH\to -\frac{\MW}{g_2\MH^2}\tr{\hV_\mu\hV^\mu}
-\frac{g_2}{2\MW\MH^2} \hpb\left(\hU\Mf\omp+\Mf\hUd\omm\right)\hp
+\OH{-4}.
\label{eq:heomexf}
\eeq
Applying this to
the complete effective Lagrangian
(i.e.\ to the bosonic and to the fermionic part), we finally find
\beqar
\delta\lreff&
=\fac\Bigg\{&\spc
\frac{g_2^2}{4\MW^2}I_{011}
\hpb\left(\hU\Mf^3\omp+\Mf^3\hUd\omm\right)\hp\nl
+\frac{ig_2^2}{8\MW^2}\left(I_{011}-2I_{112}\right)
\left[\hpb\left(\Mf^2\Slash{\hat{D}}_{f,+}\omp +
\hU\Mf^2\hUd\Slash{\hat{D}}_{f,-}\omm
\right) \hp+{\rm h.c.}\right]\nl
+\frac{i g_2^2}{8\MW^2}(I_{011}-6I_{112})
\hpb\Big[2\Mf\hUd\Slash{\hat{V}}\hU\Mf\omp-
\nl\qquad\qquad\qquad\qquad
\left(
\hU\Mf^2\hUd\Slash{\hat{V}}+\Slash{\hat{V}}\hU\Mf^2\hUd\right)\omm\Big]\hp\nl
+\frac{g_2^2}{\MW^2}\left(\frac{3}{8}I_{020}
+\frac{1}{4}I_{112}+\frac{3}{4}I_{121}
-\frac{9}{16}B_0(\MH^2,\MH,\MH)\right.\nl\qquad\qquad\!\!\left.{}
-\frac{3}{16}
{\mathrm{Re}}\left\{B_0(\MH^2,0,0)\right\}\right)
\hpb\left(
\hU\Mf\omp+\Mf\hUd\omm\right)\hp\tr{\hV_\mu\hV^\mu}\nl
+\frac{g_2^2}{2\MW^2}\left(I_{011}-2I_{112}\right)
\hpb\left[ \left(\hD_W^\mu \hV_\mu\right)\hU \Mf\omp- \Mf\hUd\left(\hD_W^\mu
\hV_\mu\right) \omm\right]\hp\nl
-\frac{g_2^4}{32\MW^4}I_{011}\left[\hpb\left(\hU
\tau_i \Mf \omp- \Mf \tau_i \hUd \omm\right) \hp\right]
\nl\qquad\qquad\qquad\qquad\times
\left[\hpbp\left(\hU
\tau_i \Mfp \omp- \Mfp \tau_i \hUd \omm\right) \hpp\right]\nl
+\frac{g_2^4}{8\MW^4}\bigg(-\frac{1}{4\MH^2}I_{010}+\frac{1}{\MH^2}
I^W_{111}(1)+\frac{9}{8}I_{020}
-\frac{9}{8}B_0(\MH^2,\MH,\MH)
\nl\qquad\qquad{}-\frac{3}{8}
{\mathrm{Re}}\left\{B_0(\MH^2,0,0)\right)\bigg\}
\left[\hpb\left(\hU\Mf\omp+\Mf\hUd\omm\right)\hp\right]
\nl\qquad\qquad\qquad\qquad\times
\left[\hpbp\left(\hU\Mfp\omp+\Mfp\hUd\omm\right)\hpp\right]\Bigg\}\nl
\back+\OH{-2}.
\label{eq:dlreff1}
\eeqar
With the  explicit
expressions for the integrals \refeq{integrals} this becomes
\beqar
\delta\lreff&=\fac\Bigg\{&\spc\frac{1}{4}
\left(\tde+1\right)\frac{g_2^2}{\MW^2}
\hpb\left(\hU\Mf^3\omp+\Mf^3\hUd\omm\right)\hp\nl
+\frac{1}{16}\left(\tde+\frac{1}{2}\right)\frac{ig_2^2}{\MW^2}
\left[\hpb\left(\Mf^2\Slash{\hat{D}}_{f,+}\omp +
\hU\Mf^2\hUd\Slash{\hat{D}}_{f,-}\omm
\right) \hp+{\rm h.c.}\right]\nl
-\frac{1}{16}\left(\tde+\frac{5}{2}\right)\frac{ig_2^2}{\MW^2}
\hpb\Big[2\Mf\hUd\Slash{\hat{V}}\hU\Mf\omp-
\nl\qquad\qquad\qquad\qquad{}
\left(
\hU\Mf^2\hUd\Slash{\hat{V}}+\Slash{\hat{V}}\hU\Mf^2\hUd\right)\omm\Big]\hp\nl
-\frac{1}{8}\left(\tde+\frac{21}{2}-\frac{3\sqrt{3}\pi}{2}\right)
\frac{g_2^2}{\MW^2}
\hpb\left(
\hU\Mf\omp +\Mf\hUd\omm
\right)\hp\tr{\hV_\mu\hV^\mu}\nl
+\frac{1}{4}\left(\tde+\frac{1}{2}\right)\frac{g_2^2}{\MW^2}
\hpb\left[ \left(\hD_W^\mu \hV_\mu\right)\hU \Mf\omp- \Mf\hUd\left(\hD_W^\mu
\hV_\mu\right) \omm\right]\hp\nl
-\frac{1}{32}\left(\tde+1\right)\frac{g_2^4}{\MW^4}
\left[\hpb\left(\hU
\tau_i \Mf \omp- \Mf \tau_i \hUd \omm\right) \hp\right]
\nl\qquad\qquad\qquad\qquad\times
\left[\hpbp\left(\hU
\tau_i \Mfp \omp- \Mfp \tau_i \hUd \omm\right) \hpp\right]\nl
-\frac{3}{64}\left(\tde+\frac{23}{3}-{\sqrt{3}\pi}\right)
\frac{g_2^4}{\MW^4}
\left[\hp\left(\hU\Mf\omp+\Mf\hUd\omm\right)\hpb\right]
\nl\qquad\qquad\qquad\qquad\times
\left[\hpp\left(\hU\Mfp\omp+\Mfp\hUd\omm\right)\hpbp\right]\Bigg\}\nl
\back+\OH{-2}.
\label{eq:dlreff2}
\eeqar

\subsection{Equations of motion and S-matrix}
\label{ssec:eom}

The tree-level and one-loop Lagrangian of the SM for $\MH\to\infty$
are given by \refeq{eq:ltree} and \refeq{eq:lfinal}, respectively.
The fermionic part of the GNLSM Lagrangian
is derived from the SM Lagrangian
\refeq{eq:lagfermi} by dropping the Higgs field in the non-linear
parametrization \refeq{eq:phinl}:
\beq
\lag_{{\mathrm{GNLSM}},F}=
i\left(\overline{\Psi}_f\Slash{D}_{f,\sigma}\om_\sigma \Psi_f
\right)
-\left(\overline{\Psi}_f\Mf U^\dagger\om_-\Psi_f
+\overline{\Psi}_f
U\Mf\om_{+}\Psi_f\right).
\label{eq:GNLSMfermi}
\eeq

The first term in \refeq{eq:dlreff2} has the same structure as the
Yukawa term in the GNLSM Lagrangian \refeq{eq:GNLSMfermi}.
Since the masses of the fermion doublet
are renormalized independently, this term can be absorbed into the
corresponding counterterm, and thus it does not contribute to the
S-matrix.

Next, we consider the second line in \refeq{eq:dlreff2} which is related
to the kinetic term in \refeq{eq:GNLSMfermi}. The
$\omp$-part can be
completely absorbed into
the counterterm to the kinetic terms for
the right-handed fermion fields since these are renormalized
independently. For the $\omm$-part
it is useful to decompose $\Mf$ \refeq{eq:doublet} as
\cite{apetal}
\beq
\Mf= \frac{1}{2}(m_{\Pf_1}+m_{\Pf_2}){\bf 1}+
\frac{1}{2}(m_{\Pf_1}-m_{\Pf_2})\tau_3
\label{eq:Mfdecomp}
\eeq
and $\Mf^2$ accordingly.
The contribution proportional to the unit matrix
inserted into the $\omm$-term yields a term, which can
be absorbed into the kinetic term of the left-handed fermion doublet.
Thus, the only
part of the second line in \refeq{eq:dlreff2} which contributes to the S-Matrix
is
\beq
\left.\delta\lreff(\mbox{S-Matrix})\right|_{\sSlash{\hD}\,\hp}=\fac
\frac{1}{32}
\left(\tde+\frac{1}{2}\right)\frac{ig_2^2}{\MW^2}\left(m_{\Pf_1}^2-
m_{\Pf_2}^2\right)\left[\hpb\hT\Slash{\hD}_{f,-}\omm\hp+{\rm h.c.}\right],
\eeq
with $\hT$ defined in \refeq{eq:V}.

Finally, we may use the classical
EOMs for the background fields in order to remove the
$\hD_W^\mu\hV_\mu$-terms in $\lreff$. Such an application of the EOM
within the effective interaction term corresponds to a shift of the
background fields which does not effect S-matrix elements \cite{eom}.
Relation \refeq{eq:DVEOM} was derived for the pure bosonic sector
of the SM. Taking into account
massive fermions, the EOM for the SU(2)$_\PW$ gauge
fields within the GNLSM become
\beq
\DW^\mu \hW_{\mu\nu} = -\frac{i}{g_2}\MW^2\hV_\nu +
PA_{1,\nu}\quad \mbox{with} \quad A_{1,\nu}^{ab}=-\frac{g_2}{2}
\hpb^b\gamma_\nu\omm\hp^a,
\label{eq:eom0}
\eeq
and  \refeq{eq:DVEOM2} generalizes to
\beq
\quad \DW^\mu \hV_\mu=PA_2\quad\mbox{with}\quad A_2^{ab}=
\frac{ig_2^2}{2\MW^2}\left[
\left(\overline{\Slash{\hat{D}}_{f,-}\omm\hp}\right)^b\hp^a
+\hpb^b\left(\Slash{\hat{D}}_{f,-}\omm\hp\right)^a\right],
\label{eq:eom1}
\eeq
where $P$ is the operator defined in \refeq{eq:P}.
In \refeq{eq:eom0} and
\refeq{eq:eom1} and the following, the indices $a$ and $b$ denote the
SU(2)$_\PW$ indices of the 2$\times$2-matrices $A_i$.
Then, we can apply the EOMs for the fermion fields
within the GNLSM
\beq
\Slash{\hat{D}}_{f,-}\omm\hp= -i\hU\Mf\omp\hp,\qquad
\overline{\Slash{\hat{D}}_{f,-}\omm\hp}=i\hpb\omm\Mf\hUd,
\eeq
and find
\beq
\DW^\mu \hV_\mu=P A_3 \quad\mbox{with}\quad A_3^{ab}=
\frac{g_2^2}{2\MW^2}\left[
\hpb^b\omp\left(\hU\Mf\hp\right)^a-\left(\hpb\Mf\hUd\right)^b\omm\hp^a\right].
\label{eq:DVEOMf}
\eeq
Applying this to the $\DW^\mu \hV_\mu$-term in \refeq{eq:dlreff2}
one finds
\beqar
&&\hspace{-2em}
\hpb\left[ \left(\hD_W^\mu \hV_\mu\right)\hU \Mf\omp- \Mf\hUd\left(\hD_W^\mu
\hV_\mu\right)\omm\right]\hp\nn\\&=&\frac{g_2^2}{4\MW^2}
\left[\hpb\left(\hU
\tau_i \Mf \omp- \Mf \tau_i \hUd \omm\right) \hp\right]
\left[\hpbp\left(\hU
\tau_i \Mfp \omp- \Mfp \tau_i \hUd \omm\right) \hpp\right],
\hspace{2em}
\label{eq:umform1}
\eeqar
and inserting this into $\lag_{11}$ of \refeq{eq:sttr},
one obtains
\beqar
\lag_{11}=\frac{g_2^4}{8\MW^4}
\left[\hpb\left(\hU
\tau_i \Mf \omp- \Mf \tau_i \hUd \omm\right) \hp\right]
\left[\hpbp\left(\hU
\tau_i \Mfp \omp- \Mfp \tau_i \hUd \omm\right) \hpp\right]. &&
\hspace{2em}
\label{eq:umform2}
\eeqar
To derive \refeq{eq:umform1} and \refeq{eq:umform2}, we have used
the definition \refeq{eq:P} and the identity
\beq
\tr{(PAU)(PBU)}=\tr{(PUA)(PUB)}
\label{eq:Pind}
\eeq
where $A$ and $B$ are arbitrary 2$\times$2-matrices
and $U$ is an SU(2) matrix.
Equation \refeq{eq:Pind} is proven in \refapp{app:Pind}.
Thus, if one considers massive fermions, the contribution of
$\lag_{11}$ to S-matrix elements does not vanish unlike in the pure
bosonic sector. $\lag_{11}$ yields an effective four-fermion
interaction which is quartic in the fermion masses.
With \refeq{eq:umform1} and \refeq{eq:umform2} the
$\DW^\mu \hV_\mu$-terms in
\refeq{eq:lefffinalex}
and \refeq{eq:dlreff2} take the form of one of the four-fermion terms
already present in \refeq{eq:dlreff2}.

Considering renormalization and the use of the EOMs,
the fermionic contribution to the Lagrangian $\lreff$(S-matrix)
\refeq{eq:finalresult}, which contains all effects of the heavy Higgs
boson on S-matrix elements, is given by%
\footnote{Note that in the linear parametrization of the SM no
$(\hpb\tau_i\hp)^2$-
and $\hpb(\hD_\PW\hV)\hp$-terms are generated directly,
because they correspond to diagrams with
$\hpb\hp\vp H$-couplings, which only exist in the non-linear
parametrization. Thus, within that framework
the only contribution to the $(\hpb\tau_i\hp)^2$-term
comes from $\lag_{11}$ according to \refeq{eq:umform2}.
Applying \refeq{eq:umform2} to the $\lag_{11}$-term in
\citere{hemo}, where the linear parametrization was used, we find that our
result for the $(\hpb\tau_i\hp)^2$-term
is consistent with the one of
\citere{hemo}; i.e.\ the difference in the $\lag_{11}$-term
between \citere{hemo} and this article is compensated by fermionic terms.}
\beqar
\rlap{$\delta\lreff(\mbox{S-matrix})=\fac\Bigg\{$}&&\nl
\phantom{+}\,\,\,
\frac{1}{32}
\left(\tde+\frac{1}{2}\right)\frac{ig_2^2}{\MW^2}\left(m_{\Pf_1}^2-
m_{\Pf_2}^2\right)\left[\hpb\hT \Slash{\hD}_{f,-}\omm\hp+{\rm h.c.}\right]\nl
-\frac{1}{16}\left(\tde+\frac{5}{2}\right)\frac{ig_2^2}{\MW^2}
\hpb\Big[2\Mf\hUd\Slash{\hat{V}}\hU\Mf\omp-
\left(
\hU\Mf^2\hUd\Slash{\hat{V}}+\Slash{\hat{V}}\hU\Mf^2\hUd\right)\omm\Big]\hp\nl
-\frac{1}{8}\left(\tde+\frac{21}{2}-\frac{3\sqrt{3}\pi}{2}\right)
\frac{g_2^2}{\MW^2}
\hpb\left(\hU\Mf\omp+\Mf\hUd\omm \right)\hp\tr{\hV_\mu\hV^\mu}\nl
-\frac{1}{192}\frac{g_2^4}{\MW^4}\left[\hpb\left(\hU
\tau_i \Mf \omp- \Mf \tau_i \hUd \omm\right) \hp\right]
\left[\hpbp\left(\hU
\tau_i \Mfp \omp- \Mfp \tau_i \hUd \omm\right) \hpp\right]\nl
-\frac{3}{64}\left(\tde+\frac{23}{3}-{\sqrt{3}\pi}\right)
\frac{g_2^4}{\MW^4}
\left[\hpb\left(\hU\Mf\omp+\Mf\hUd\omm\right)\hp\right]
\nl\qquad\qquad\qquad\qquad\times
\left[\hpbp\left(\hU\Mfp\omp+\Mfp\hUd\omm\right)\hpp\right]\Bigg\}\nl
\!\!\!+\OH{-2}.
\label{eq:finalresultfermi}
\eeqar

\section{Discussion of the result}
\label{sec:dis}

Inspecting
the bosonic part of the effective Lagrangian
\refeq{eq:finalresult}, we see that the
first two terms contribute to vector-boson two-point (and higher)
functions, the third and the fourth
to vector-boson three-point (and higher) functions, and the
last two to vector-boson four-point functions. This means that the first two
terms parametrize the effects of the heavy Higgs boson on
LEP~1 physics, the next two become relevant for
LEP~2 physics, and the last two for LHC physics.

By naive power counting one expects that
integrating out the Higgs boson generates dimension-2 terms at
$\OH{2}$ and dimension-4 terms at $\OH{0}$ (i.e.\ proportional to
$\log\MH$) \cite{hemo,apbe,long}. Actually, only those effective
terms which do not violate custodial SU(2)$_\PW$ invariance are generated
at this order. However, the effective Lagrangian
\refeq{eq:lefffinalex} contains only one custodial-SU(2)$_\PW$-violating
term%
\footnote{Strictly speaking, the designation ``custodial SU(2)$_\PW$
invariance'', i.e.\ global SU(2)$_\PW$ invariance in the absence of the
$B$-field, is misleading, because locally
SU(2)$_\PW\times$U(1)$_{\mathrm Y}$-invariant
terms as in \refeq{eq:sttr} automatically fulfill this
invariance. In the literature  the expression
``custodial-SU(2)$_\PW$-invariant'' is commonly
used for terms which are custodial-SU(2)$_\PW$-invariant
when additionally the Goldstone fields are disregarded
(rhs of \refeq{eq:sttr}),
and in this sense it also has to be
understood in this article.
The custodial-SU(2)$_\PW$-violating terms are then those containing the
operator $\hT$ \refeq{eq:V} but not explicitly the $\hB$-field.},
namely $\lag_0$ \refeq{eq:sttr}.
This is a dimension-2 term;
nevertheless it is only generated at $\OH{0}$. There are 7
custodial-SU(2)$_\PW$-violating
dimension-4 terms \cite{hemo,long} but none of
them
is generated at $\OH{0}$. This means that custodial-SU(2)$_\PW$-breaking
terms are suppressed by
at least a factor
of $\MW^2/\MH^2$ in comparison to the
prediction of naive power counting. Actually, the reason for this
suppression also follows from a (slightly more involved) power
counting argument: The custodial-SU(2)$_\PW$-breaking terms are those
which explicitly
contain the operator $P_3$ defined in \refeq{eq:P}. However,
as shown in \refse{sec:helim}, all contributions from that operator
to $\tilde{\tilde{\Delta}}_{\ss H}\dpp$ and thus to $\leff$ have the form
$(\MW^2-\MZ^2)P_3$ (see eq.~\refeq{eq:P3}).
Therefore, $P_3$ always occurs together with a power of $\MW^2$ and
for dimensional reasons these contributions are suppressed by
an additional power of $\MW^2/\MH^2$.

The fermionic part  of the effective Lagrangian
\refeq{eq:finalresultfermi} contains contributions to
fermion two-point functions in the first term, to
fermion-fermion-vector couplings in the first and the second term,
fermion-fermion-vector-vector couplings in the third term and
four-fermion interactions in the last two terms.
All effective fermionic couplings have
at least a factor $m_{\Pf_i}/\MW$.
Consequently, the fermionic part of the effective Lagrangian
\refeq{eq:finalresultfermi} vanishes for massless fermions
(and is suppressed for light fermions),
i.e.\ the purely bosonic effective Lagrangian \refeq{eq:finalresult}
describes all $\OH{0}$-effects of the heavy Higgs boson in this case.
Unlike the bosonic terms, the effective fermionic interactions
of course break custodial SU(2)$_\PW$ owing to the occurrence of the
non-degenerate fermion-mass matrix $\Mf$
\refeq{eq:doublet}. Furthermore,
also effective fermionic terms of dimension 5 or 6
are generated at $\OH{0}$
and not only dimension-4 terms like in the bosonic sector.

In analogy to the simpler SU(2) toy model considered in \citere{sdcgk},
we find that the limit $\MH\to\infty$ of the standard model
at one loop is the corresponding GNLSM plus the effective
interaction terms given in \refeq{eq:finalresult} and
\refeq{eq:finalresultfermi}, which describe the one-loop
effects of the heavy Higgs boson. In order to calculate the complete
one-loop effects to a given process at $\OH{0}$,
one still has to consider the
effects of the light quantum fields in the GNLSM Lagrangian.
The coefficients of the effective terms in \refeq{eq:finalresult},
\refeq{eq:finalresultfermi}
contain logarithmic divergences $\Delta$ (see \refeq{tde}).
Since the SM is is renormalizable, these UV-divergences necessarily
cancel against the logarithmically divergent contributions of the
non-renormalizable one-loop Lagrangian of the GNLSM
$\lag_{\mathrm{GNLSM}}^{\mathrm{1-loop}}$ in \refeq{eq:lfinal}.
These have been calculated for the bosonic part of the GNLSM
in \citere{long} and for the dimension-4 terms of the fermionic part in
\citere{apetal}. Comparing our result \refeq{eq:finalresult} with
\citere{long} and the first two terms in \refeq{eq:finalresultfermi}
with
\citere{apetal}\footnote{In order to compare \refeq{eq:finalresultfermi}
with \citere{apetal} one has to decompose $\Mf$ \refeq{eq:doublet}
according to \refeq{eq:Mfdecomp}.
The logarithmically divergent contributions of the GNLSM to the
fermionic dimension-5 and -6 terms (third to fifth term in
\refeq{eq:finalresultfermi}) have to our knowledge not been calculated
in the literature.}
we find that the divergencies indeed cancel.
In particular, since logarithmic divergences and
$\log\MH$-terms always occur in the linear combination $\tde$
\refeq{tde}, the logarithmically divergent one-loop
contributions of the GNLSM to S-matrix elements coincide with the
logarithmically $\MH$-dependent one-loop contributions in the SM, if
one replaces
\beq
\frac{2}{4-D}-\gamma_E+\log(4\pi)+\log\mu^2 \quad\to\quad \log\MH^2.
\label{eq:subs}
\eeq
However, the Lagrangians \refeq{eq:finalresult}
and \refeq{eq:finalresultfermi} contain additional finite and
$\MH$-independent contributions. Thus, the $\log\MH$ one-loop
contributions to
the S-matrix in the SM can alternatively be calculated in the GNLSM
with the replacement \refeq{eq:subs}, however the constant
contribution cannot be calculated within this model.
Therefore, the GNLSM is {\em not} identical to the limit $\MH\to\infty$
of the SM beyond tree-level.
In this context, it should be kept in mind that these results are
derived in dimensional regularization.

The non-decoupling one-loop contributions of a heavy Higgs boson to
physical observables can directly be read
from the effective Lagrangians \refeq{eq:finalresult}
and \refeq{eq:finalresultfermi} simply by
calculating the contributions of the
generated effective terms (which only contain background
fields) at  tree level.

\section{Physical applications}
\label{sec:examples}

In this section
we illustrate the use of the constructed
effective Lagrangian.
We derive
the heavy-Higgs effects
for some vertex functions and
transition amplitudes
directly from our effective Lagrangian. As a consistency check, we
compare the results with those of a diagrammatical calculation.

We skip the well-known heavy-Higgs effects on LEP1 observables, where
the Higgs-boson dependence is merely due to vacuum-polarization effects
in the gauge-boson propagators. The corresponding $\log\MH$-terms can
easily be read off from the
first two lines in the
effective Lagrangian \refeq{eq:finalresult}.

\subsection{Bosonic processes}

We start by considering vector-boson scattering. In \citere{aaww} the
heavy-Higgs effects on the one-loop radiative corrections to
$\gamma\gamma\to\PWp\PWm$ in the SM have been investigated and related
to the corrections within the GNLSM. From our Lagrangian
\refeq{eq:finalresult} it is very easy to reproduce the results given
there so that we do not repeat the explicit formulas. We just note that
no $\log\MH$-terms in the SM with a heavy Higgs boson appear, i.e.\ the
one-loop corrections to $\gamma\gamma\to\PWp\PWm$ in the GNLSM are
UV-finite despite of the non-renormalizability of the GNLSM.

As a second example we treat the process
$$ \PWp(k_1,\lambda_1) + \PWm(k_2,\lambda_2) \to
   \PWp(k_3,\lambda_3) + \PWm(k_4,\lambda_4) $$
in the heavy-Higgs limit. Here $k_{1,2}$ denote the (incoming) momenta
of the incoming W bosons,
and $k_{3,4}$ the (outgoing) momenta of the
outgoing W bosons. The corresponding Mandelstam variables are defined by
\beq
s = (k_1+k_2)^2, \quad
t = (k_1-k_3)^2, \quad
u = (k_1-k_4)^2.
\eeq
The helicity states are labeled by $\lambda_i$, and the corresponding
polarization vectors by $\varepsilon_i$.
In the limit $s,-t,-u,\MW^2\ll\MH^2$
the tree-level transition amplitude ${\cal M}_0$ is given by
\beqar
{\cal M}_0 &=&
\frac{4\pi\alpha}{\sw^2}\left[
\frac{{\cal M}_s}{s-\MZ^2}
+(\varepsilon_1\cdot\varepsilon^*_4)(\varepsilon_2\cdot\varepsilon^*_3)
-(\varepsilon_1\cdot\varepsilon_2)(\varepsilon^*_3\cdot\varepsilon^*_4)
\right]
-4\pi\alpha \frac{\MZ^2}{s(s-\MZ^2)}{\cal M}_s
\nn\\[.3em] &&{}
\;+\; \mbox{crossed}
\;+\; \OH{-2} ,
\label{eq:4Wborn}
\eeqar
where crossing means the interchanges
$\varepsilon_2\leftrightarrow\varepsilon^*_3, k_2\leftrightarrow -k_3$.
Note that the single contributions in \refeq{eq:4Wborn} are arranged
according to the independent couplings $g_2=e/\sw$ and $e$, where
$\alpha=e^2/4\pi$ is the usual fine-structure constant.
The following shorthands have been introduced,
\beqar
{\cal M}_s &=& {\cal M}^\prime_s
+(u-t)(\varepsilon_1\cdot\varepsilon_2)(\varepsilon^*_3\cdot\varepsilon^*_4)
\nn\\ &&{}
+2(\varepsilon_1\cdot\varepsilon_2)\left[
    (k_1\cdot\varepsilon^*_4)(k_2\cdot\varepsilon^*_3)
   -(k_1\cdot\varepsilon^*_3)(k_2\cdot\varepsilon^*_4)\right]
\nn\\ &&{}
+2(\varepsilon^*_3\cdot\varepsilon^*_4)\left[
    (k_3\cdot\varepsilon_2)(k_4\cdot\varepsilon_1)
   -(k_3\cdot\varepsilon_1)(k_4\cdot\varepsilon_2)\right],
\nn\\[.5em]
{\cal M}^\prime_s &=& \phantom{{}+{}}
4(k_1\cdot\varepsilon_2)\left[
    (k_3\cdot\varepsilon^*_4)(\varepsilon_1\cdot\varepsilon^*_3)
   -(k_4\cdot\varepsilon^*_3)(\varepsilon_1\cdot\varepsilon^*_4)\right]
\nn\\ &&{}
+4(k_2\cdot\varepsilon_1)\left[
    (k_4\cdot\varepsilon^*_3)(\varepsilon_2\cdot\varepsilon^*_4)
   -(k_3\cdot\varepsilon^*_4)(\varepsilon_2\cdot\varepsilon^*_3)\right]
\nn\\ &&{}
+2(\varepsilon_1\cdot\varepsilon_2)\left[
    (k_1\cdot\varepsilon^*_4)(k_2\cdot\varepsilon^*_3)
   -(k_1\cdot\varepsilon^*_3)(k_2\cdot\varepsilon^*_4)\right]
\nn\\ &&{}
+2(\varepsilon^*_3\cdot\varepsilon^*_4)\left[
    (k_3\cdot\varepsilon_2)(k_4\cdot\varepsilon_1)
   -(k_3\cdot\varepsilon_1)(k_4\cdot\varepsilon_2)\right].
\eeqar
Now, we consider the one-loop effects of the heavy Higgs boson to this
process, which can be obtained from the
effective Lagrangians \refeq{eq:lefffinalex} or
\refeq{eq:finalresult}, respectively,
simply by calculating the tree-level contributions of $\lreff$.
As explained above, only the terms
in \refeq{eq:finalresult} are relevant for the contribution to the
S-matrix element, whereas the additional terms in
\refeq{eq:lefffinalex}
cancel exactly. The effective Lagrangian yields the difference
$\de{\cal M}=\de{\cal M}_{\mathrm{SM}}-\de{\cal M}_{\mathrm{GNLSM}}$ (in
dimensional regularization) between the
one-loop corrections to the amplitude in the SM with a heavy Higgs boson
and the
GNLSM, respectively. One finds
\beqar
\de{\cal M} &=&
\spc\frac{\alpha^2}{\sw^4}\Biggl[
{}-\frac{5}{6}\left(\tde+\frac{19}{30}\right)
\left(\frac{{\cal M}_s}{s-\MZ^2}
+(\varepsilon_1\cdot\varepsilon^*_4)(\varepsilon_2\cdot\varepsilon^*_3)
-(\varepsilon_1\cdot\varepsilon_2)(\varepsilon^*_3\cdot\varepsilon^*_4)
\right)
\nn\\[.3em] &&{}
\phantom{{}+\frac{\alpha^2}{\sw^4}\Biggl[}
-\frac{1}{12}\left(\tde+\frac{17}{6}\right)
(\varepsilon_1\cdot\varepsilon^*_4)(\varepsilon_2\cdot\varepsilon^*_3)
\nn\\[.3em] &&{}
\phantom{{}+\frac{\alpha^2}{\sw^4}\Biggl[}
-\frac{1}{6}\left(\tde+\frac{175}{12}-\frac{9\sqrt{3}\pi}{4}\right)
(\varepsilon_1\cdot\varepsilon_2)(\varepsilon^*_3\cdot\varepsilon^*_4)
\Biggr]
\nn\\[.3em] &&{}
-\frac{\alpha^2}{\sw^2}\frac{1}{6}
\frac{\MZ^2}{s(s-\MZ^2)}{\cal M}^\prime_s
\quad \;+\; \mbox{crossed}
\;+\; \OH{-2} .
\label{eq:4WHH}
\eeqar
The single terms in \refeq{eq:4WHH} are arranged such that only the
second and the third line
yield contributions of order $xy/\MW^4$ ($x,y=s,t,u$)
in the high-energy
limit for purely longitudinally polarized W bosons. These terms entirely
originate from the genuine four-point operators in the effective
Lagrangian, i.e.\ from $\lag_4$ and $\lag_5$.
The complete $xy/\MW^4$-terms of the one-loop correction to
$\PW^+_{\mathrm L}\PW^-_{\mathrm L}\to\PW^+_{\mathrm L}\PW^-_{\mathrm L}$
in the limit $\MW^2\ll s,-t,-u\ll\MH^2$ were
calculated in \citere{veyn} and \citere{dawi} in an SU(2) gauge theory
and the SM, respectively.
Comparing our results with the ones given there, we find
agreement for the $\log\MH$-terms
\footnote{The terms of the order $(xy/\MW^4)\log\MH$ were already given
in \citere{chey} by calculating the logarithmic divergences
($\Delta$-terms) within the GNLSM and using the replacement \refeq{eq:subs}.}
and the ``$\sqrt{3}\pi$'' term, which
stems from Higgs-mass renormalization. The remaining $\MH$-independent
$xy/\MW^4$-terms are of course different since additional terms of this kind
originate from bosonic loops without Higgs bosons, which are equal in
the SM and GNLSM.
As a consistency check, we have also calculated $\delta{\cal M}$
diagrammatically and found the same result.
\begin{figure}
\begin{center}
\begin{picture}(12,2.5)
\put(-3.8,-15.0){\includegraphics{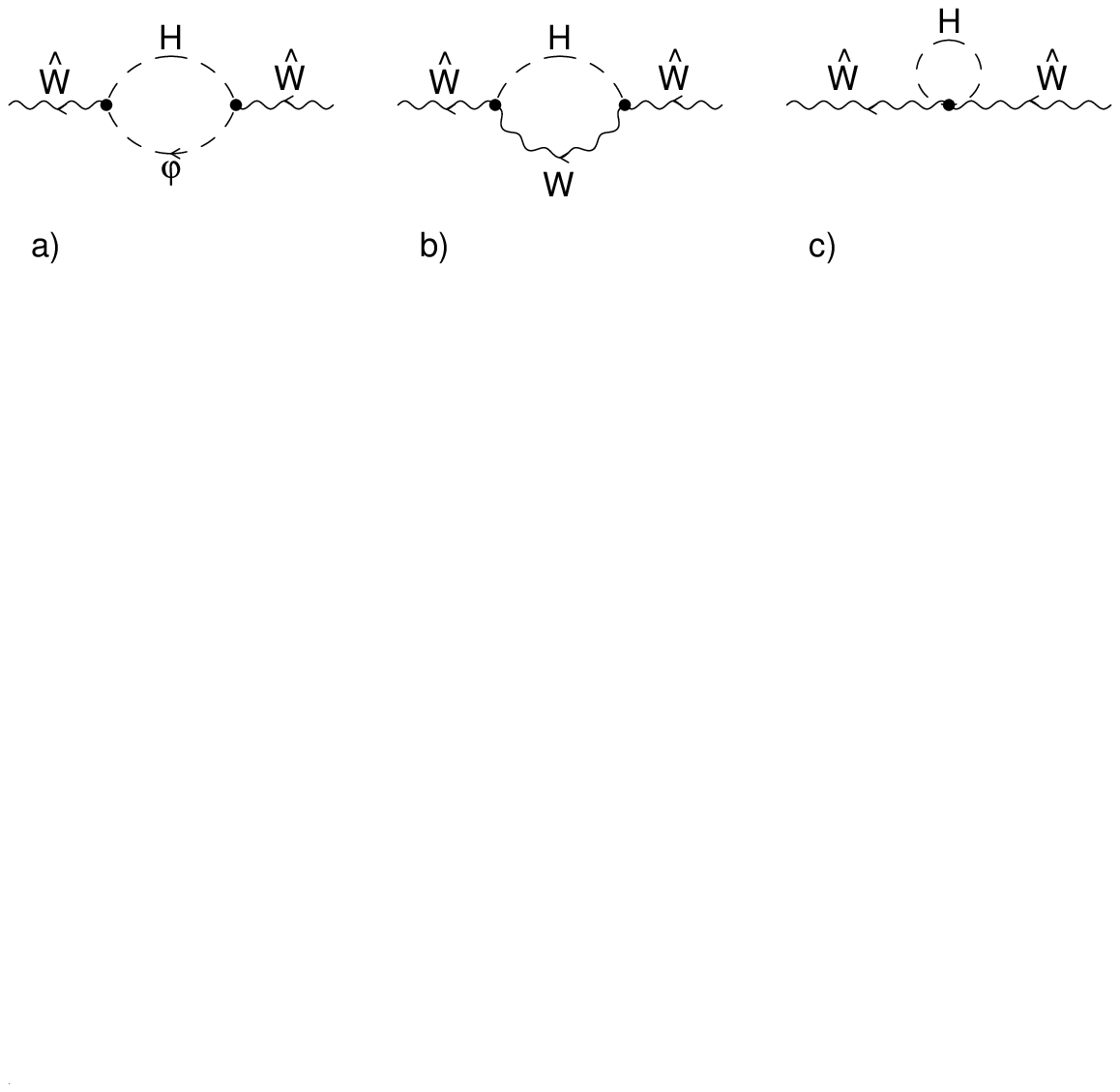}}
\end{picture}
\end{center}
\caption{Higgs diagrams to the $\hW$--self-energy.}
\label{WW}
\begin{center}
\begin{picture}(12,2.5)
\put(-3.8,-15.0){\includegraphics{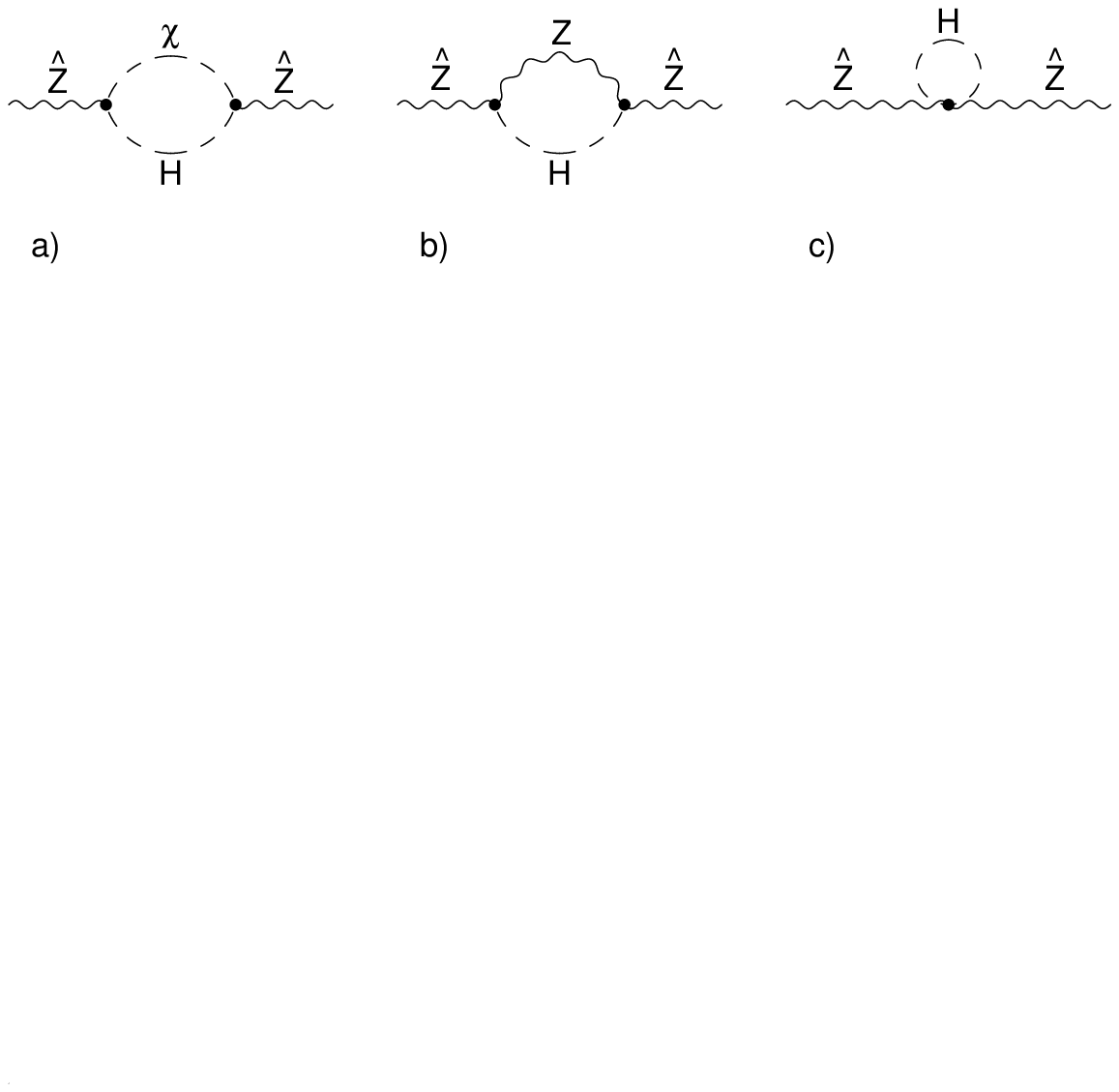}}
\end{picture}
\end{center}
\caption{Higgs diagrams to the $\hZ$--self-energy.}
\label{ZZ}
\begin{center}
\begin{picture}(12,5.5)
\put(-3.8,-12.0){\includegraphics{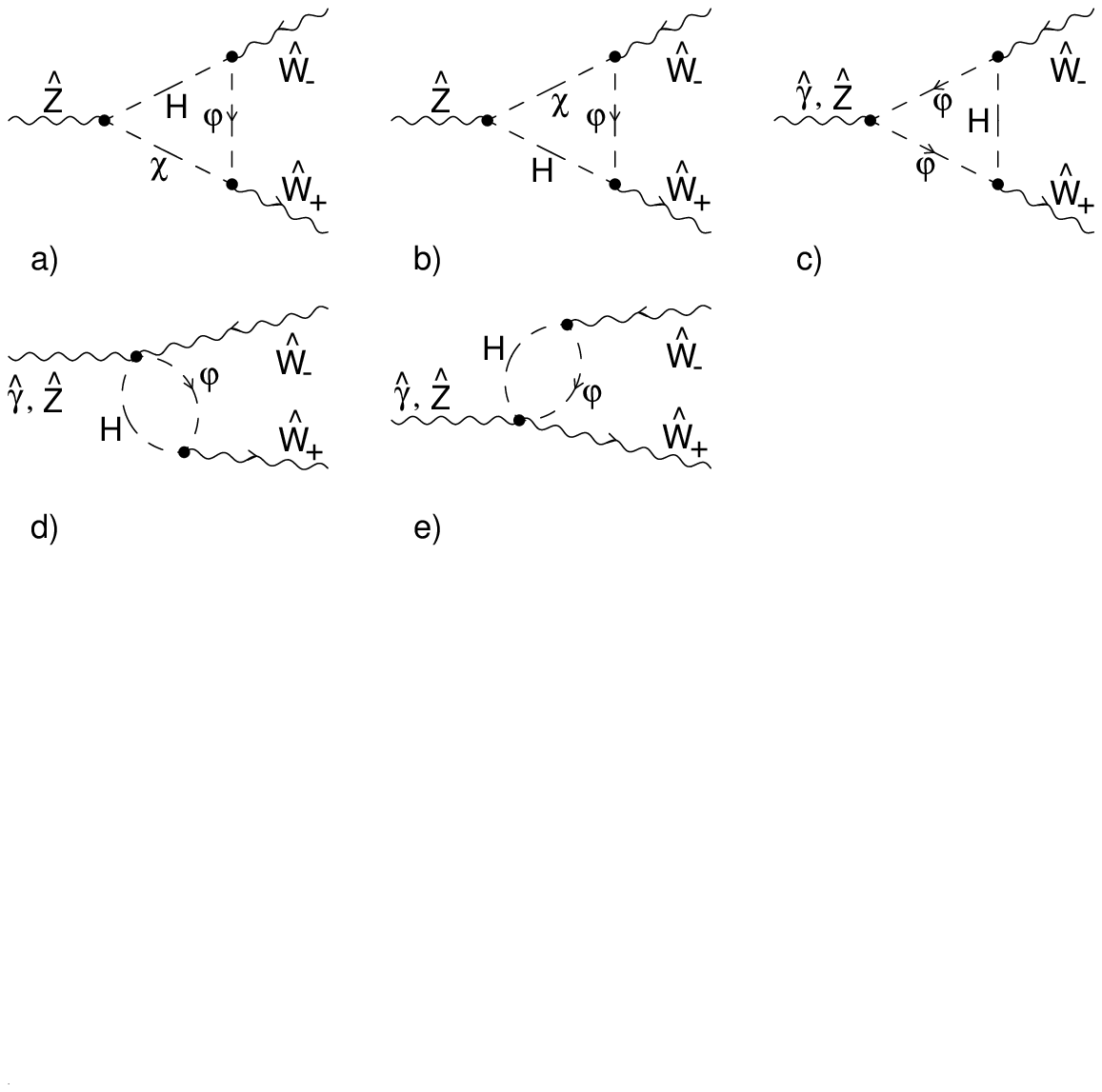}}
\end{picture}
\end{center}
\caption{Higgs diagrams of $\OH{0}$ for the $\hZ\hW^+\hW^-$- and
$\hA\hW^+\hW^-$-vertex functions.}
\label{ZWWAWW}
\efi
\begin{figure}
\begin{center}
\begin{picture}(16,17.0)
\put(-3.1,-5.3){\includegraphics{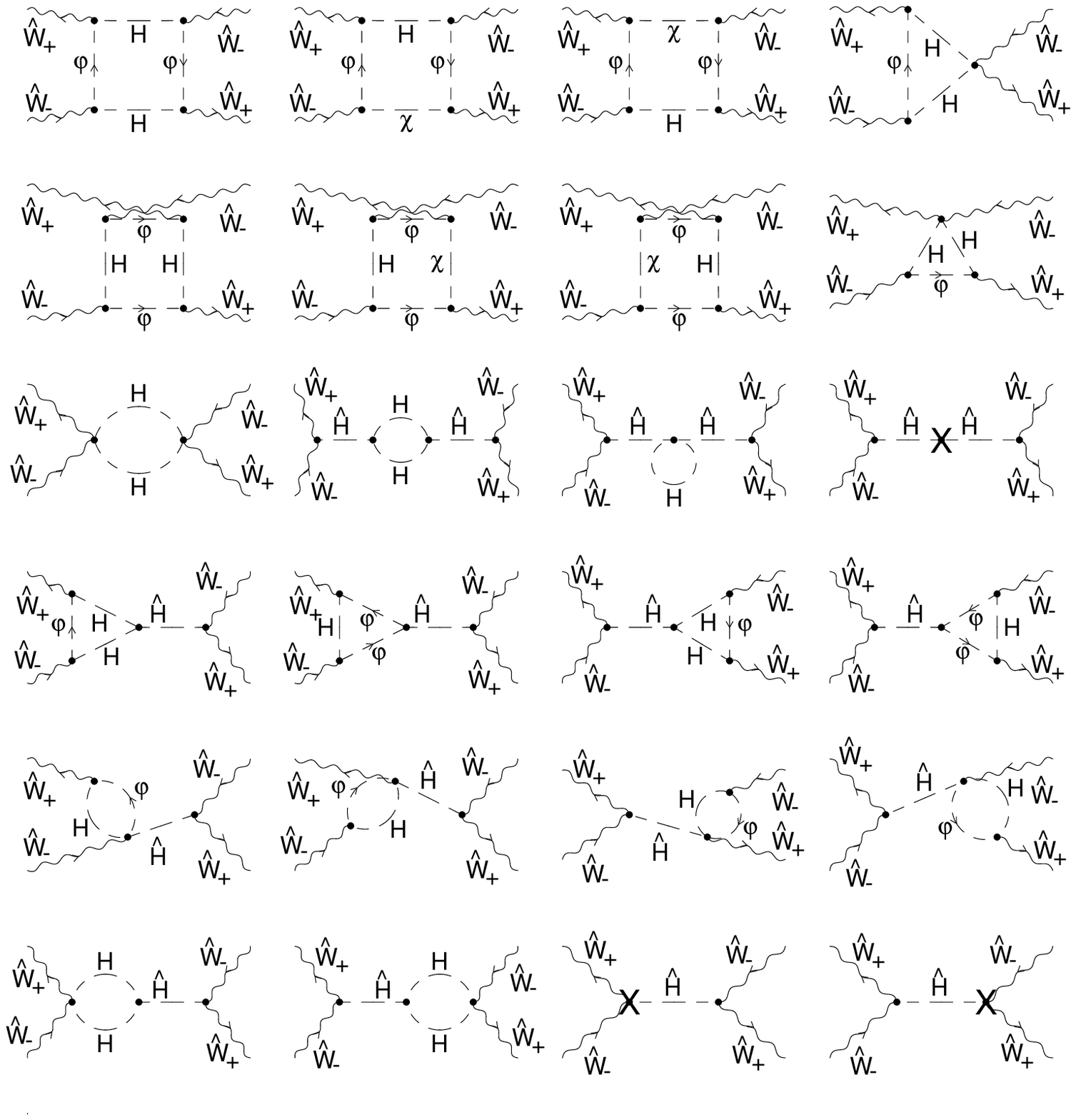}}
\put(3,0){\mbox{+ crossed graphs (external $\hW^-$ interchanged)}}
\end{picture}
\end{center}
\caption{Higgs diagrams of $\OH{0}$ for the one-particle-irreducible
and the heavy-Higgs reducible
$\hW^+\hW^-\hW^+\hW^-$--four-point function.}
\label{4Wpm}
\efi
Figures \ref{WW},\ref{ZZ},\ref{ZWWAWW} show the Higgs-mass-dependent
subdiagrams contributing in $\OH{0}$ to Feynman diagrams and
counterterms which are reducible with respect to light particles.
The irreducible $\OH{0}$ contributions and those which are
reducible with respect to the heavy Higgs field
(which correspond to the irreducible contributions of $\lreff$)
are depicted in \reffi{4Wpm} (where all fields are
assumed to be incoming).
The advantage of our effective-Lagrangian approach is obvious:
in a diagrammatical calculation all these diagrams
have to be evaluated while in the effective-Lagrangian
calculation one only has to consider the tree-level contributions of
$\lreff$ \refeq{eq:finalresult}.

\subsection{Fermionic processes}

Now, we turn to examples involving massive fermions.
The only Higgs-mass-dependent
contributions of the effective Lagrangian \refeq{eq:dlreff1}
to the
fermion self-energy are contained in the first two terms, viz.\
\begin{figure}
\begin{center}
\begin{picture}(12,3)
\put(-3.8,-14.8){\includegraphics{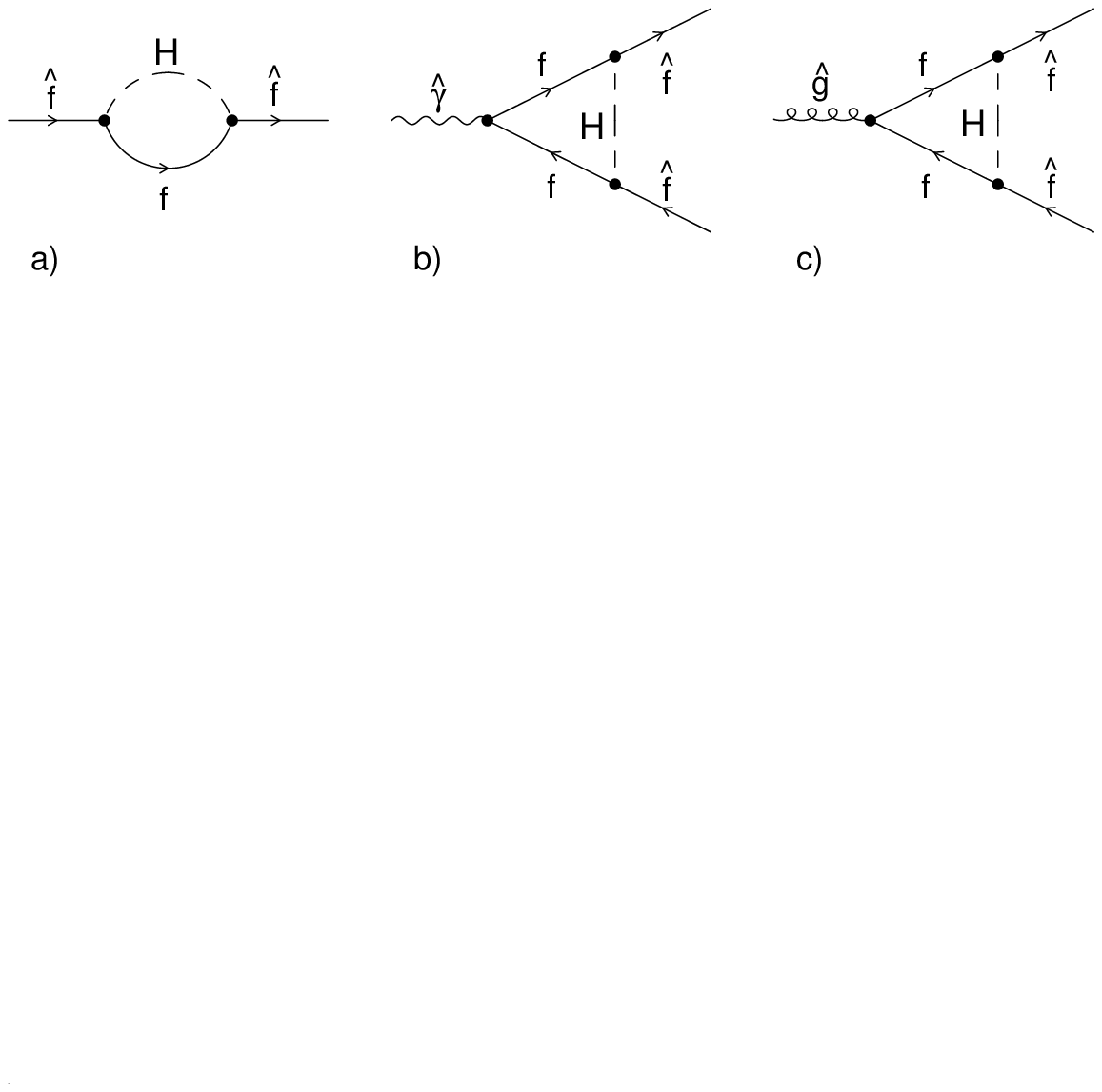}}
\end{picture}
\end{center}
\caption{Higgs diagrams contributing to the a) fermion self-energy, b)
photon-fermion-fermion vertex, c) gluon-fermion-fermion vertex.}
\label{Hff}
\efi
\beqar
\delta\Si^{\hfibar\hfi}_{\mathrm L}(k^2) &=&
\delta\Si^{\hfibar\hfi}_{\mathrm R}(k^2) =
\frac{g_2^2}{64\pi^2}\frac{\mfi^2}{\MW^2}\left(I_{011}-2I_{112}\right)
\;+\;\OH{-2}, \nn\\
\delta\Si^{\hfibar\hfi}_{\mathrm S}(k^2) &=&
\frac{g_2^2}{64\pi^2}\frac{\mfi^2}{\MW^2}I_{011}
\;+\;\OH{-2},
\label{eq:ffse}
\eeqar
where our conventions for the fermionic self-energy follow the ones of
\citere{bfm5}.
In a diagrammatical calculation, these contributions stem from
the graph of \reffi{Hff}.a).
Using \refeq{eq:ffse}, we get for the contributions to the
renormalization constants,
\beqar
\left.\frac{\de\mfi}{\mfi}\right|_{\mathrm H} &=&
\frac{g_2^2}{32\pi^2}\frac{\mfi^2}{\MW^2}\left(I_{011}-2I_{112}\right)
\;+\;\OH{-2}, \nn\\
\left.\de Z_{f_i}^{\si}\right|_{\mathrm H} &=&
-\frac{g_2^2}{64\pi^2}\frac{\mfi^2}{\MW^2}\left(I_{011}-2I_{112}\right)
\;+\;\OH{-2}.
\label{eq:ffcts}
\eeqar
The field-renormalization constants
$\de Z_{f_i}^{\si}$ are chosen such
that the residue of the $f_i$ propagator equals one. Combining
\refeq{eq:ffse} and \refeq{eq:ffcts}, we obtain that the renormalized
fermion self-energy contains no Higgs-mass-dependent terms of $\OH{0}$,
\beq
\delta\Si^{\hfibar\hfi,{\mathrm{ren}}}_{\mathrm{L/R/S}}(k^2) \;=\; \OH{-2}.
\label{eq:ffseren}
\eeq
The Higgs-mass dependence of the photon-fermion-fermion vertex is
contained in the second term in \refeq{eq:dlreff1}, which yields
\beq
\de\Ga^{\hA\hfibar\hfi}_\mu(k,\bar p,p) =
-\frac{iQ_{\Pf_i} eg_2^2}{64\pi^2}\frac{\mfi^2}{\MW^2}\ga_\mu
\left(I_{011}-2I_{112}\right) \;+\; \OH{-2}.
\label{eq:Aff}
\eeq
In a diagrammatical calculation one has to calculate
the graph shown in \reffi{Hff}.b).
Again after renormalization no $\OH{0}$
survives for this vertex function,
\beq
\de\Ga^{\hA\hfibar\hfi,{\mathrm{ren}}}_\mu(k,\bar p,p) \;=\; \OH{-2}.
\label{eq:Affren}
\eeq
The $\OH{0}$ contributions to $\de\Ga^{\hA\hfibar\hfi}_\mu$ are cancelled
by the fermionic wave-function corrections, and the charge
renormalization constant does not contain terms of $\OH{0}$.
{}From \refeq{eq:ffseren} and \refeq{eq:Affren} we draw the conclusion
that no $\OH{0}$-terms of the effective Lagrangian contribute
e.g.\ to the SM one-loop corrections to $\gamma\gamma\to\Pf_i\Pfbar_i$. This
means that the SM one-loop
prediction for $\gamma\gamma\to\Pf_i\Pfbar_i$ in the
heavy-Higgs limit approaches asymptotically the GNLSM correction, which is
UV-finite either. The analogue conclusion also holds for gluon-gluon fusion,
$\Pg\Pg\to\Pf_i\Pfbar_i$, since the Higgs-mass-dependent subdiagrams of
$\OH{0}$ are the same as for $\gamma\gamma\to\Pf_i\Pfbar_i$ with the
external photons replaced by gluons. More precisely, only the diagrams
shown in \reffis{Hff}a),c) are relevant. For instance,
the complete SM one-loop correction
to $\Pg\Pg\to\Pt\Ptbar$ can be found in \citere{ggtt}. From the results
given there, one can see that the relative one-loop correction approaches
a constant for $\MH\to\infty$ in consistence with our result.

The result \refeq{eq:Affren} is in agreement with the one obtained in
\citere{mayu} for the $\ga tt$-vertex.
Inspecting our corresponding results for the fermion-mass-dependent terms
of the $ttZ$- and the $tbW$-vertices,
\beqar
\left.\delta\Gamma^{\mathrm{\hZ\hat{\bar{t}}
\hat{t},ren}}_\mu(k,\bar{p},p)\right|_{\Mf}&=&
\frac{ig_2^3}{128\pi^2\cw}\frac{\Mt^2}{\MW^2}
\gamma_\mu\gamma_5
\left(I_{011}-6I_{112}\right)+\mbox{($k_\mu$-terms)}+\OH{-2},\nn\\
\left.\delta\Gamma^{\mathrm{\hW\hat{\bar{t}}\hat{b},
ren}}_\mu(k,\bar{p},p)\right|_{\Mf}&=&
-\frac{ig_2^3}{128\sqrt{2}\pi^2}\gamma_\mu\left(\frac{\Mt^2+\Mb^2}{\MW^2}\omm
-2\frac{\Mt\Mb}{\MW^2}\omp\right)
\left(I_{011}-6I_{112}\right)\nl+\mbox{($k_\mu$-terms)}+\OH{-2}.
\label{eq:tbW}
\eeqar
which are contained in the second and
third terms in \refeq{eq:dlreff1}%
\footnote{As indicated in \refeq{eq:tbW}, there are also $k_\mu$-terms
stemming from the fifth term in \refeq{eq:dlreff1}.
As explained in \refsse{ssec:eom}, this term
becomes a four-fermion term in $\lreff$(S-matrix)
\refeq{eq:finalresultfermi} after applying the EOM.
Thus, its contribution is not considered here.},
we also find agreement with \citere{mayu}, where the
$\Mt^2\log\MH$-terms were calculated.

Finally,
we investigate the heavy-Higgs effects to the
top-quark decay $\Pt\to\PWp\Pb$.
In lowest order the transition amplitude for this process is given by
\beq
{\cal M}_0 = \frac{e}{\sqrt{2}\sw}
\bar{u}(p_\Pb)\Slash{\varepsilon}^*_\PW\omega_- u(p_\Pt),
\eeq
with $p_\Pt$ and $u(p_\Pt)$ ($p_\Pb$ and $u(p_\Pb)$) denoting the incoming
(outgoing) momentum and spinor for the top(bottom)-quark, respectively.
$\varepsilon_\PW$ represents the polarization vector of the W boson.
The complete
difference $\de{\cal M}=\de{\cal M}_{\mathrm{SM}}-\de{\cal
M}_{\mathrm{GNLSM}}$
can easily be calculated from the effective interaction
terms
\refeq{eq:tbW}. We obtain
\beqar
\de{\cal M} &=& \frac{e\alpha}{16\sqrt{2}\pi\sw^3}\Biggl\{
\phantom{{}+{}}
\bar{u}(p_\Pb)\Slash{\varepsilon}^*_\PW\omega_- u(p_\Pt)
\left[\left(\frac{\Mt^2+\Mb^2}{\MW^2}\right)
\frac{1}{4}\left(\tde+\frac{5}{2}\right)
-\frac{11}{6}\left(\tde+\frac{5}{6}\right)
\right]
\nn\\ && {}
\phantom{\frac{e\alpha}{16\sqrt{2}\pi\sw^3}\Biggl\{}
-\bar{u}(p_\Pb)\Slash{\varepsilon}^*_\PW\omega_+ u(p_\Pt)
\frac{\Mt\Mb}{\MW^2}\frac{1}{2}\left(\tde+\frac{5}{2}\right)
\Biggr\}
\;+\;\OH{-2}.
\label{eq:tbWHH}
\eeqar
\begin{figure}
\begin{center}
\begin{picture}(16,3)
\put(-3.8,-14.8){\includegraphics{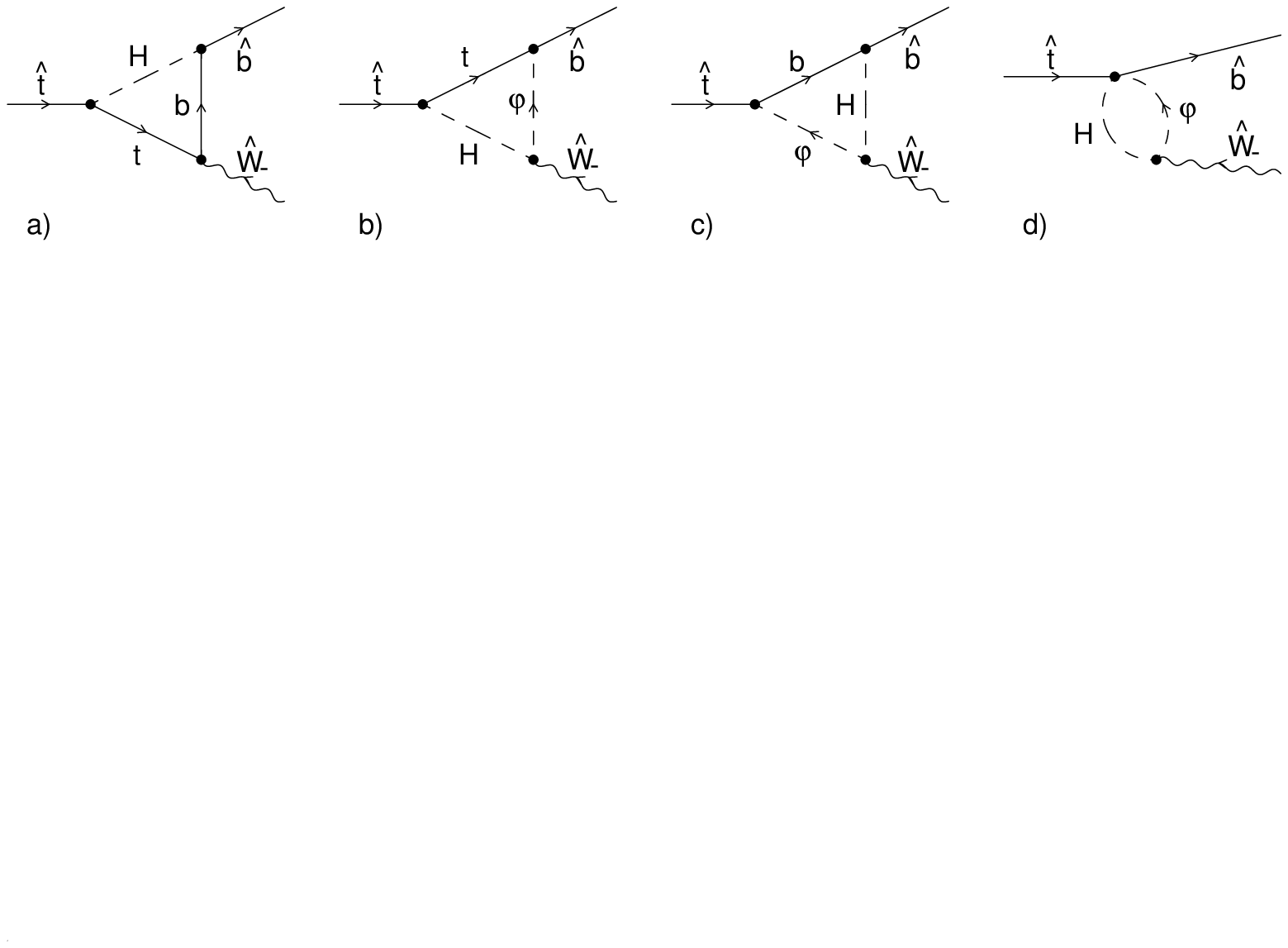}}
\end{picture}
\end{center}
\caption{Higgs diagrams of $\OH{0}$ for the
$\hat{t}\hat{\bar b}\hW^-$-vertex function.}
\label{tbW}
\efi
Alternatively, \refeq{eq:tbWHH} could be derived by calculating the
diagrams shown in \reffi{tbW}, where graph \ref{tbW}.d) does not
contribute to the S-matrix element.
The term in \refeq{eq:tbWHH} which is not multiplied by fermion masses
is entirely due to coupling-constant and W-wave-function
renormalization. It is associated with the well-known variable
$\Delta r$, i.e.\ it is absent in a renormalization scheme, where the
Fermi constant $G_F$ is used as an input parameter instead of the W mass
$\MW$. The $\MH$ dependence of the top width originating from the
remaining $\log\MH$-terms in \refeq{eq:tbWHH} is e.g.\ numerically
discussed in \citere{topwidth1}, where the complete one-loop SM
correction is calculated. The $(\Mt^2/\MW^2)\log\MH$-term can for
instance be found
in \citere{topwidth2} in agreement with our result.

\section{Conclusion}
\label{sec:sum}
In this article we
have
integrated out the Higgs boson in the
electroweak standard model
directly in the path integral, assuming that it is very heavy.
We have expressed all non-decoupling
effects, i.e.\ effects of $\OH{0}$,
of the heavy Higgs boson
(including fermionic effects)
in terms of an effective Lagrangian,
from which the leading contributions of the Higgs boson
to physical parameters and scattering processes can easily be read.

For the bosonic sector of the SM, this
result itself is essentially already known from the
diagrammatical calculation of \citere{hemo}. However, we have derived
it in a completely different way, viz.\ by integrating out the Higgs
boson directly in the path integral instead of calculating Feynman
diagrams and matching the full theory to the effective one.
The functional method is a methodical progress for several
reasons: As pointed out in \citere{esma}, diagrammatical calculations
like those
in \citere{hemo} cannot determine the full content of Green function
but only the  ``physically relevant parts''.
This is due to problems with
gauge invariance
of the matching conditions. However, owing to the
application of the background-field method and the Stueckelberg formalism,
our direct calculation yields the complete effective Lagrangian
in a manifestly gauge-invariant form without those problems.
Moreover, the functional method is a huge technical simplification in
comparison to the diagrammatical one, because in the functional
approach the effective Lagrangian -- which contains contributions to
many Green functions -- is generated
{\em directly} by integrating out the heavy field.
In a diagrammatical calculation one has to calculate various Green
functions (i.e.\ very many Feynman graphs),
to write down all effective interaction terms which
could possibly be generated, and then determine the
effective Lagrangian by comparing coefficients \cite{hemo}. We can
use the convenient matrix notation throughout, i.e.\ we do not have
to specify the single components of the fields. For the background
fields we even do not have to introduce the  physical basis.
A striking simplification within our
method is the fact that it is completely
obvious that only 7 of 14 possible effective bosonic
interaction terms of dimension 4 (or 2) are generated in
$\OH{0}$ at one loop, i.e.\ that the 7 custodial-SU(2)$_\PW$-violating
dimension-4 terms are only of $\OH{-2}$. This result was also found by
the diagrammatical calculation in \citere{hemo}, however
no obvious reason why these terms cancel can be seen there.
In our direct calculation these terms are not generated
from the beginning; i.e.\
there are no cancellations. The suppression of all
custodial-SU(2)$_\PW$-violating terms by one power of $\MW^2/\MH^2$
follows in our approach from a simple power-counting argument.

In addition, we also considered the fermionic sector of the standard
model when integrating out the Higgs field, and constructed the
fermionic terms of the effective Lagrangian. These have not been
completely
calculated before, neither functionally nor diagrammatically. Also
this calculation becomes straightforward owing to the use of our
functional method. If one applied the diagrammatical method, one
would have to
write down all possible effective interaction terms in order to
find the matching conditions. Since even dimension-5 and -6 terms are
are generated, this would be a large number, while in a functional
calculation also these terms are generated directly.

In the present article we have integrated out a {\em non-decoupling\/} heavy
field.
However, the generalization of our method to the case of {\em decoupling\/}
fields is straightforward yielding a wide field of
phenomenologically interesting applications.

\section*{Acknowledgement}
C.G.-K.\ thanks the University of Bielefeld for hospitality during his
visit.

\appendix
\def\theequation{\thesection.\arabic{equation}}
\setcounter{equation}{0}
\section*{Appendix}

\section{Explicit expressions for the one-loop integrals}
\label{app:ints}

In \refse{sec:helim} the construction of the unrenormalized effective
Lagrangian (\ref{eq:leff1}) was traced back to the vacuum integrals
$I^i_{klm}(\xi)$ defined in (\ref{eq:intnot}). Such vacuum integrals are
easily calculated
and their explicit expressions are
already given in the appendix of
\citere{sdcgk} using dimensional regularization.
The relevant $\OH{0}$ parts of the $I^i_{klm}$ for $D\to 4$ are
\beqar
I_{010}\phantom{(\xi)}&=&\MH^2(\tde+1),\nn\\
I_{011}^i(\xi)&=&\tde+1 +\OH{-2},\nn\\
I_{020}\phantom{(\xi)}&=&\tde,\nn\\
I_{111}^i(\xi)&=&\frac{1}{4}(\MH^2+\xi M_i^2)
             \left(\tde+\frac{3}{2}\right)+\OH{-2},\nn\\
I_{112}^i(\xi)&=&\frac{1}{4}\left(\tde+\frac{3}{2}\right)+\OH{-2},\nn\\
I_{121}^i(\xi)&=&\frac{1}{4}\left(\tde+\frac{1}{2}\right)+\OH{-2},\nn\\
I_{213}^i(\xi)&=&\frac{1}{24}\left(\tde+\frac{11}{6}\right)+\OH{-2},\nn\\
I_{222}^i(\xi)&=&\frac{1}{24}\left(\tde+\frac{5}{6}\right)+\OH{-2}
\label{integrals}
\eeqar
with
\beq
\tde=\De-\log\left(\frac{\MH^2}{\mu^2}\right), \qquad
\De =\frac{2}{4-D}-\gamma_E+\log(4\pi),
\label{tde}\eeq
and $\gamma_E$ being Euler's constant.
In the main part of this article we
drop the index $i$ and the
argument $\xi$  for all logarithmically divergent integrals, because
these are independent of $M_i^2$ and $\xi$ at $\OH{0}$.

In \refse{sec:ren} we expressed the renormalization constant $\de\MH^2$
(\ref{eq:dmh}) in terms of the
$I_{klm}$ and scalar two-point functions
$B_0(k^2,M_1,M_2)$ defined in (\ref{eq:b0}). The explicit expressions for
the relevant $B_0$-functions can for instance be deduced from the general
result presented in \citere{de93}, leading to
\beqar
B_0(\MH^2,\MH,\MH) &=& \tde+2-\frac{\pi}{\sqrt{3}},
\nn\\
B_0(\MH^2,0,0)     &=& \tde+2+i\pi.
\label{Bs}\eeqar

\section{Proof of equation \protect\refeq{eq:Pind}}
\label{app:Pind}

In this appendix we prove
relation \refeq{eq:Pind}, which has been
used in order to simplify the $\DW^\mu \hV_\mu$-terms in $\lreff$ by
using the
EOMs.

First, we derive the identity
\beq
P \left(U A U^\dagger \right)= U (PA) U^\dagger,
\label{eq:id1}
\eeq
where $P$ is the projection operator \refeq{eq:P}, $A$ an
arbitrary 2$\times$2-matrix and $U$ an SU(2) matrix.
Using the definition of $P$ we find
\beq
P\left(U AU^\dagger\right)=\frac{1}{2}\tau_i\tr{\tau_iU AU^\dagger}
=\frac{1}{2} \tau_i \tr{U^\dagger\tau_iU A}.
\label{eq:id2}
\eeq
Owing to
$\tr{U^\dagger\tau_i U}=\tr{\tau_i}=0$,
the hermitian 2$\times$2-matrix
$U^\dagger\tau_iU$ is a linear combination of Pauli matrices,
i.e.\ it can be written as
\beq
U^\dagger\tau_iU  = X_{ij}\tau_j
\qquad\mbox{with}\qquad
X_{ij}=\frac{1}{2}\tr{\tau_j U^\dagger \tau_i U}.
\label{eq:id3}
\eeq
This implies
\beq
U \tau_j U^\dagger= \tau_i X_{ij}.
\label{eq:id4}
\eeq
With \refeq{eq:id2}, \refeq{eq:id3}, \refeq{eq:id4} and \refeq{eq:P}
we find
\beq
P\left(U AU^\dagger\right)=\frac{1}{2}\tau_iX_{ij}\tr{\tau_j A}
=\frac{1}{2}U\tau_j U^\dagger \tr{\tau_j A}
=U (PA) U^\dagger,
\eeq
which proves \refeq{eq:id1}. With \refeq{eq:id1} one can easily derive
\refeq{eq:Pind}:
\beqar
\tr{(PAU)(PBU)} &=&
\tr{U (PAU) U^\dagger U (PBU) U^\dagger} = \tr{(PUA)(PUB)}.
\eeqar

\end{document}